\documentclass{article}

\usepackage[version=3]{mhchem} 
\usepackage{siunitx} 
\usepackage[margin=1in]{geometry}
\usepackage{xcolor}
\usepackage{enumitem}
\usepackage{moresize}
\usepackage{graphicx} 

\usepackage{amsmath} 
\usepackage{physics}
\usepackage{mathtools}
\usepackage{amssymb}
\usepackage{ntheorem}
\usepackage{feynmp}
\usepackage{tikz}
\usepackage{pgfplots}

\DeclareMathOperator{\Hol}{Hol}
\DeclareMathOperator{\vol}{vol}

\newtheorem*{definition}{Def$^{\textnormal{\textbf{n}}}$}
\newtheorem*{thm}{Th$^{\textnormal{\textbf{m}}}$}

\begin{document}

\baselineskip=18pt  
\numberwithin{equation}{section}  
\allowdisplaybreaks  


\thispagestyle{empty}

\vspace*{1.8cm}
\begin{center} {\Huge G$_{2}$-Manifolds and\\
    \vspace{.2cm} M-Theory Compactifications}

\vspace*{1.5cm}
Aaron Kennon
\vspace*{1.0cm}

{\it Department of Physics, \\
University of California, Santa Barbara, CA 93106, USA}\\
{\tt akennon@physics.ucsb.edu} \\ \vspace{.5cm}

\vspace*{0.8cm}
\end{center}
\vspace*{1cm}

The mathematical features of a string theory compactification determine the physics of the effective four-dimensional theory. For this reason, understanding the mathematical structure of the possible compactification spaces is of profound importance. It is well established that the compactification space for M-Theory must be a seven-manifold with holonomy $G_{2}$, but much else remains to be understood regarding how to achieve a physically-realistic effective theory from such a compactification. Much also remains unknown about the mathematics of these $G_{2}$-Manifolds, as they are quite difficult to construct. This review discusses progress with regards to both the mathematical and physical considerations surrounding spaces of holonomy $G_{2}$. Special attention is given to the known constructions of $G_{2}$-Manifolds and the physics of their corresponding M-Theory compactifications. 

\newpage

\clearpage
\tableofcontents


\newpage

\section{$G_{2}$-Manifolds}

One of the most outlandish claims of string theory is that there necessarily exist additional spacetime dimensions besides the four of everyday experience. One possible explanation for why we don't experience these dimensions is that they are compactified. Direct detections of these extra dimensions are presumably impossible at observable energy scales; however, they would in principle affect measurements of the four-dimensional effective theory. More concretely, the mathematical structure of the manifold of extra dimensions would completely determine the physical predictions of the theory. The entire spacetime manifold then may be expressed locally as the product of a four-dimensional non-compact Lorentzian Manifold with a Riemannian Manifold of appropriate dimension. 
\begin{equation} 
M=M_{4} \times M_{d}
\end{equation}

In the case of the heterotic superstring, consistency with $\mathcal{N}=1$ supersymmetry for the effective theory requires the six-dimensional (compact) space corresponding to the extra dimensions to be a Calabi-Yau three-fold. A significant amount of effort has gone into studying these spaces from both mathematical \cite{ghj, Yau1} and physical perspectives \cite{Hubsch, Davies}. Techniques from complex and algebraic geometry are quite useful for the study of Calabi-Yau manifolds, particularly with regards to their construction. One standard way of constructing Calabi-Yau n-folds is to start with n+1 dimensional complex projective space and consider submanifolds whose canonical bundle is trivial (a defining condition for a compact K{\"a}hler Manifold to be Calabi-Yau). The adjunction formula of algebraic geometry provides the condition for the the canonical bundle of these submanifolds to vanish \cite{Hartshorne}. For instance, such a construction in the three-dimensional case produces the quintic three-fold, famous in the context of mirror symmetry and enumerative geometry \cite{bbs}. \\

\subsection{Motivating Physical Interest in $G_{2}$-Manifolds}

In contrast to superstring theories which are ten-dimensional, M-Theory is an eleven-dimensional theory and so its compactification space is seven-dimensional. As a consequence, the mathematical structure of the M-Theory compactification space is necessarily distinct from that of the superstring compactifications. However, analogously to the superstring, the physical criterion for determining the mathematical structure of the space centers around achieving $\mathcal{N}=1$ supersymmetry for the effective four-dimensional theory. The reason for expecting this degree of supersymmetry follows from the necessary requirement that if the effective four-dimensional theory has any supersymmetry, it must be broken at some energy scale in order to make contact with the phenomenology of the standard model.  In particular, there exists a family of $\mathcal{N}=1$ supersymmetric extensions to the standard model that are realistic and do break supersymmetry at an appropriate energy scale \cite{bbs}. The extensions with $\mathcal{N}\geq 2$ are not realistic because such models suggest that massless fermions always transform in a real representation of the gauge group, which is incompatible with what we observe \cite{gsw}. However, we do want some amount of supersymmetry to make the theory phenomenologically appealing. By this reasoning, we can assume that the effective theory has $\mathcal{N}=1$ supersymmetry. We also expect to find the compactification space to be Ricci-Flat, since this property ensures that the spacetime manifold written in the form of equation 1.1 satisfies the eleven-dimensional vacuum Einstein Equations. \\

These physical considerations translate directly into a more tangible mathematical constraint on the compactification space \cite{Witten1}. Unbroken supersymmetries imply that the vacuum expectation values of the fields are unchanged under supersymmetry transformations, so we can effectively count the residual supersymmetries by considering the number of fields whose vacuum expectation values are unchanged under these transformations. The Bose fields are automatically invariant under such transformations, so the amount of supersymmetry in the effective theory is determined by the change in the vacuum expectation values of the fermi fields. This change is captured by the change in the gravitino field $\psi$ \cite{Duff}.  
\begin{equation}
\delta\psi_{m}=\nabla_{m} \xi-\left( \frac{i}{144}\Gamma^{abcd}_{m}+8\Gamma^{bcd}\delta^{a}_{m}\right) G_{abcd} \xi
\end{equation}

Here $\xi$ is an eleven-dimensional spinor and the factors of $\Gamma$ correspond to the eleven-dimensional Dirac matrices. In the case that the gravitino field is invariant under the supersymmetry transformation, and if the G-Flux vanishes, the gravitino equation simplifies to the standard Killing Spinor Equation. 
\begin{equation}
\nabla \xi=0
\end{equation}

This equation suggests that there must exist a non-trivial covariantly constant spinor on spacetime. Noting the compactification ansatz, we can write this spinor as the tensor product of a spinor acting on M$_{4}$ and another acting on M$_{7}$. Requiring $\mathcal{N}=1$ supersymmetry in the effective theory then directly implies that there should also be exactly one covariantly constant spinor on the compactification space. \\

This correspondence is incredibly useful because it translates the physical requirement of $\mathcal{N}=1$ supersymmetry into the mathematical condition of existence of a covariantly constant spinor. Ultimately though, the most convenient way to characterize $\mathcal{N}=1$ supersymmetry is not in terms of the spinor directly, but through its implications for the holonomy group of the manifold under the spin connection. 
\begin{definition} Holonomy Group of M under connection $\nabla$. \textnormal{Let (M, g) be an n-dimensional Riemannian Manifold, let E be a vector bundle over (M, g), and let $\nabla$ be a connection on E. Fix a base point x $\in$ M and consider the set of all parallel transport maps along the fibres $p:E_{x}\rightarrow E_{x}$ corresponding to closed loops based at x. This collection of maps forms a group Hol(M, x) called the \textit{Holonomy Group of M under the connection $\nabla$} based at x $\in$ M.} 
\end{definition} 

Note from the definition that the holonomy group depends not only on the choice of manifold and base point, but also on the connection. In our case the vector bundle will be the spin bundle and the connection the spin connection, but oftentimes the holonomy group is defined in terms of parallel transport maps of vectors under the Levi-Civita connection. Note that in this latter case, since parallel transport maps preserve the lengths of vectors, the holonomy group is automatically a subgroup of O(n), and if the manifold is oriented it is a subgroup of SO(n). By analogous reasoning, the spin holonomy group is automatically a subgroup of Spin(n). In the literature the term \textit{restricted holonomy} is commonly used to refer to a manifold whose holonomy group is a proper subgroup of SO(n) or Spin(n) in the cases of the Levi-Civita or spin connections, respectively. Also note that if the manifold is path-connected then the holonomy groups based at various points on the manifold will be isomorphic, so we can ignore the base point in discussions of the holonomy group. The spaces will be presumed to be path-connected throughout this review. \\

We now make three additional mathematical assumptions which will ultimately allow us to uniquely specify the holonomy group corresponding to an M-Theory compactification space. Specifically, we require that the manifold be simply-connected, irreducible, and non-symmetric. 
\begin{definition} Irreducible. \textnormal{A Riemannian Manifold (M, g) is said to be \textit{irreducible} if no finite cover of the manifold is isometric to a direct product of manifolds each of strictly lesser dimension than M.}
\end{definition}

\begin{definition} Riemannian Symmetric Space. \textnormal{A Riemannian Manifold (M, g) is said to be \textit{symmetric} if for every $x\in M$ there is an isometry that is an involution and for which the point x is an isolated fixed point.}
\end{definition}

Intuitively, an irreducible manifold doesn't globally resemble a product of manifolds of lesser dimensions and a symmetric space may be thought of as a space whose symmetry group includes inversions about each point. Physically we can motivate the assumption of simple-connectedness in relation to considerations of spin structures, which we want since we are considering spinors. The necessary and sufficient condition for an oriented Riemannian manifold to admit a spin structure is for the second Stiefel-Whitney class to vanish \cite{JoyceB}. The existence of spin structures will be automatic from the spaces we are considering, but they will be unique if the space is simply-connected, which is one reason why it is a desirable quality. The irreducibility assumption follows directly from the compactification ansatz and although it is not a priori obvious, it can be shown from the classification of Riemannian Symmetric spaces that there are no simply-connected irreducible symmetric seven-manifolds that are Ricci-Flat \cite{Helgason}. As a consequence they are not of physical interest in the context of M-Theory compactifications and we can specialize to the non-symmetric case. \\

Given that we are considering spaces that are simply-connected, irreducible, and non-symmetric, we can look to the Berger Classification to determine the possible holonomy groups \cite{Berger}. In its original context, the Berger Classification specifies possible holonomy groups for simply-connected, irreducible, non-symmetric spaces under the Levi-Civita Connection, so in this form the classification is not relevant. However, based on the relationship of the spin connection to the Levi-Civita connection, we do know that the holonomy group under the spin connection is either isomorphic to the holonomy group under the Levi-Civita connection, or the homomorphism is a double cover. Using the assumption of simple-connectedness, we can rule out the latter possibility, thereby demonstrating that the Berger Classification applies equally well to the possible holonomy groups under the spin connection \cite{JoyceB}. \\
 
Looking at the possible holonomy groups corresponding to dimension seven in the Berger Classification we can see that the exceptional Lie group $G_{2}$ is the only viable candidate. Reassuringly for physicists, spaces of holonomy G$_{2}$, and even those whose holonomy is contained in G$_{2}$, are automatically Ricci-Flat \cite{JoyceB}. Following from these considerations, it is clear that not only do spaces of $G_{2}$-Holonomy have fundamental importance in mathematics as a result of their role in the Berger Classification, but they are also fundamentally important to physics as the compactification spaces for M-Theory. Ultimately though, we will see that physically-realistic compactification spaces are \textit{necessarily} singular so they cannot actually be smooth manifolds.

\subsection{$G_{2}$ and Representations}

The group $G_{2}$ occupies a fundamental place in the Cartan classification as the smallest in the subclass of exceptional Lie groups \cite{JoyceB}. The group G$_{2}$ is fourteen-dimensional and topologically it is compact, connected, and simply-connected. Its Lie Algebra $\mathfrak{g}_{2}$ is semisimple and has rank 2. The group G$_{2}$ itself may be defined in terms of the octonion algebra, the division algebra that is neither ordered, commutative, nor associative.  In fact, the groups corresponding to the Berger Classification relate quite generally to the various division algebras. To clarify this relationship, it is useful to note that the space of octonions may be decomposed into so-called real and imaginary parts as follows \cite{Baez}. 
\begin{equation}
\mathbb{O}=\mathbb{R} \oplus \mathbb{R}^7
\end{equation}

One way to define the group G$_{2}$ is in terms of these imaginary octonions. 
\begin{definition} The Exceptional Lie Group G$_{2}$. \textnormal{The group \textit{G$_{2}$} is defined to be the automorphism group of the imaginary octonions.}
\end{definition}

We may then consider a vector in the part isomorphic to $\mathbb{R}^7$ and define a vector cross product that takes two vectors in the imaginary octonions and outputs a third. In fact, the only dimensions in which we are able to define such a vector product are dimensions 3 and 7 \cite{fg}. In the latter case, the group that preserves this cross-product is $G_{2}$. As a consequence, it can be shown that $G_{2}$ is the subgroup of Spin(7) that fixes a unit vector on S$^{7}$. This characterization is immediately relevant to the covariantly constant spinor whose existence was a consequence of $\mathcal{N}=1$ supersymmetry. This interpretation directly implies the following relationship between groups. 
\begin{equation}
\textnormal{Spin}(7)/G_{2}=S^{7}
\end{equation}

The definition of G$_{2}$ in terms of the automorphism group of the imaginary octonions may be considered the most fundamental, but there is an equivalent notion that relates more directly to the theory of G$_{2}$-Holonomy. To arrive at this definition, define a three-form $\Phi_{0}$ in terms of standard coordinates of $\mathbb{R}^{7}$ in the following way. 
\begin{equation}
\Phi_{0}= dx_{123} +dx_{145} +dx_{167}+dx_{246}-dx_{257}-dx_{347}-dx_{356}
\end{equation}

Here $dx_{ijk}=dx_{i} \wedge dx_{j} \wedge dx_{k}$. Then $G_{2}$ may also be defined as the stabilizer of $\Phi_{0}$ in $\mathbb{R}^{7}$ \cite{JoyceB}. This notion of $G_{2}$ as the stabilizer of the three-form $\Phi_{0}$ is completely equivalent to $G_{2}$ as the automorphism group of the octonion algebra. This three-form used in the latter definition is central to the theory of manifolds of holonomy $G_{2}$. \\

In terms of the study of G$_{2}$-Holonomy spaces, the representations of the group G$_{2}$ are particularly important. $G_{2}$ has a seven-dimensional fundamental vector representation and a fourteen-dimensional adjoint representation. These representations may be realized by studying the Lie Algebra of $G_{2}$ in comparison to that of SO(7). The Lie Algebra of $G_{2}$ may be given in terms of the three-form $\Phi_{0}$. 
\begin{equation}
\mathfrak{g_{2}}=\{ \alpha \in \mathfrak{so}(7) : \Phi_{0abc} \alpha^{bc}=0 \}
\end{equation}

Given the three-form $\Phi_{0}$ preserved by the group G$_{2}$, define a map $f_{\Phi_{0}}$: $\mathbb{R}^{7}$ $\rightarrow$ $\mathfrak{so}(7)$ via f$_{\Phi_{0}}(v)= v \lrcorner ~\Phi_{0}$. The Lie Algebra of SO(7) may then be related to that of G$_{2}$ in the following way. 
\begin{equation}
\mathfrak{so}(7)=\mathfrak{g_{2}} \oplus f_{\Phi_{0}}(\mathbb{R}^7)
\end{equation}

In terms of this decomposition, the adjoint representation of G$_{2}$ acts on $\mathfrak{g_{2}}$, which is 14-dimensional, and the fundamental representation acts on $f_{\Phi_{0}}(\mathbb{R}^7)$, which is seven-dimensional. \\

Practically speaking, when analyzing spaces of $G_{2}$-Holonomy we will be primarily concerned with the action of the group $G_{2}$ on spaces of smooth k-forms of fixed degree, denoted $\Lambda^{k}(M)$. The previous considerations regarding the fundamental and adjoint representations imply that the spaces of k-forms decompose into irreducible representations of dimension $\ell$ , denoted $\Lambda_{l}^{k}(M)$ in the following ways \cite{JoyceB}. 
\begin{align}
&\Lambda^{1}=\Lambda_{7}^{1} \\
&\Lambda^{2}=\Lambda_{7}^{2} \oplus \Lambda_{14}^{2}\\
&\Lambda^{3}=\Lambda_{1}^{3} \oplus \Lambda_{7}^{3} \oplus \Lambda_{27}^{3}\\
&\Lambda^{4}=\Lambda_{1}^{4} \oplus \Lambda_{7}^{4} \oplus \Lambda_{27}^{4} \\
&\Lambda^{5}=\Lambda_{7}^{5} \oplus \Lambda_{14}^{5}\\
&\Lambda^{6}=\Lambda_{7}^{6}
\end{align}

Note that the spaces of k-forms and (n-k)-forms are related by the Hodge star in the standard sense. This decomposition for three-forms is particularly important to the theory of holonomy $G_{2}$, which is not surprising considering the role that the three-form $\Phi_{0}$ plays in the definition of the group. 

\subsection{$G_{2}$-Structures}

So far, we have seen that we can characterize $\mathcal{N}=1$ supersymmetry either in terms of a covariantly constant spinor or through the holonomy group of the manifold. From the general theory of holonomy groups, it is well understood that there is a relationship between the holonomy group of a manifold and an algebraic concept called a G-Structure. This relationship gives us another valuable way of characterizing the geometry of a space of holonomy $G_{2}$; however, we will see that a space admitting a G-Structure is only equivalent to the holonomy being contained in G, so by itself it is a weaker condition. A G-Structure is related to the frame bundle of a manifold, a concept which comes up naturally in differential geometry when translating between the languages of vector bundles and principle bundles. 
\begin{definition} Frame Bundle. \textnormal{Given an n-dimensional manifold M and basepoint x $\in$ M, we can define a new manifold N in terms of coordinates $(x,e_{1},\ldots,e_{n})$, where the $e_{i}$ are basis elements for the tangent space at x. Then associated to the projection onto the basepoint is a natural action of GL(n, $\mathbb{R}$) on the fibres. Then N is a principle bundle with fibre GL(n, $\mathbb{R}$) and it is called the \textit{Frame Bundle} of M.}
\end{definition}

\begin{definition} $G_{2}$-Structure. \textnormal{Given an n-manifold M and a Lie subgroup of GL(n, $\mathbb{R}$), a \textit{G-Structure} is a principle sub-bundle P of F with fibre G. A \textit{$G_{2}$-Structure} has fibre $G_{2}$.}
\end{definition}
 
This definition is rather unwieldy in a practical sense. Fortunately though, there is an easy way to characterize a $G_{2}$-Structure in terms of a particular three-form, often called the $G_{2}$ three-form. This $G_{2}$ three-form is ultimately very useful for characterizing the holonomy group of the manifold via strictly analytical means. The first step to realizing its importance is to note the relationship between the $G_{2}$ three-form and the $G_{2}$-Structure. There is a direct correspondence between oriented $G_{2}$-Structures and positive $G_{2}$ forms as described by a theorem of Joyce \cite{JoyceB}. 

\begin{definition} Positive Three-Form. \textnormal{Let (M, g) be a smooth, oriented Riemannian Manifold. A Three-Form $\Phi$ is called \textit{positive} if for every point $x \in M$ there is an oriented vector space isomorphism between the tangent space at x and $\mathbb{R}^{7}$ such that $\left. \Phi \right |_{x} = \Phi_{0}$.}
\end{definition} 

Considered in terms of components, this isomorphism maps the components of the three-form into those of $\Phi_{0}$ and also maps the metric components into those of $\mathbb{R}^7$ in standard coordinates. Having defined positive $G_{2}$ three-forms we may now state the correspondence. 
\begin{thm} $G_{2}$-Structure and Positive $G_{2}$ Three-Forms. \textnormal{Let M be an oriented seven-manifold. Then there exists a one-to-one correspondence between oriented $G_{2}$-Structures on M and positive three-forms. The positive three-form then locally determines a Riemannian Metric and a four-form.}
\end{thm}

The expression for the Riemannian Metric in terms of the three-form is as follows.
\begin{equation}
g_{ab}=(\det s)^{-1/9} s_{ab}
\end{equation}
\begin{equation}
s_{ab}=\frac{1}{144}\Phi_{amn}\Phi_{bpq}\Phi_{rst}\epsilon^{mnpqrst}
\end{equation}

The expression for the Hodge star follows in the standard sense from the metric and the action of the Hodge star on the $G_{2}$ three-form determines the aforementioned four-form, which is called the $G_{2}$ four-form. Note that although the Hodge star is a linear operator, it is non-linear when considered as a function of the metric. This non-linearity is one reason that the theory of G$_{2}$-Holonomy is so difficult to understand. \\

The $G_{2}$-Structure and $G_{2}$ three-form are evidently related, but what we really care about is the holonomy group. The $G_{2}$ three-form itself is related to the holonomy group of the manifold through its torsion and following from the discussed relationship between the G$_{2}$ three-form and G$_{2}$-Structure, this relationship ties together the theory of G$_{2}$-Structures and Holonomy. 
\begin{definition} Torsion. \textnormal{Let (M, g) be a Riemannian Manifold with Levi-Civita connection $\nabla$. Given a $G_{2}$ three-form $\Phi$, the \textit{torsion} is the quantity $\nabla \Phi$. If M is orientable and the three-form is positive then the torsion of a $G_{2}$-Structure is defined in terms of the torsion of its associated positive three-form.}
\end{definition}

 A $G_{2}$-Structure is said to be torsion-free if the torsion vanishes. In the general theory of holonomy groups it is possible to put a condition on the torsion that provides a necessary and sufficient condition for a manifold to have holonomy group that is a subgroup of some given Lie Group. \cite{JoyceB}.
\begin{thm} Holonomy and Torsion. \textnormal{Let (M, g) be an n-dimensional Riemannian Manifold and let G be a Lie Subgroup of GL(n $\mathbb{R}$). Then the manifold has holonomy contained in G if and only if it admits a torsion-free G-Structure.}
\end{thm}

As a result of this relationship, people often refer to a torsion-free $G_{2}$-Structure in terms of a positive torsion-free three-form even though formally speaking the $G_{2}$-Structure is merely determined by this three-form. \\

We have now seen that we can characterize the holonomy group of a manifold most directly in terms of a specific three-form, which itself determines a $G_{2}$-Structure. There exists one more characterization of holonomy derived from the three-form which is often useful. This condition follows from considering the torsion classes of $G_{2}$-Structures which makes use of the decomposition of the space of three-forms as presented in equation 1.11 \cite{fg}. Namely, we can decompose the torsion $\nabla \Phi$ in terms of the torsion classes in the following way. 
\begin{equation}
\nabla\phi =\tau_{1}g+\tau_{7}+\tau_{14}+\tau_{27}
\end{equation}

Based on this analysis we can express the torsion components in terms of the exterior derivatives of the $G_{2}$ three-form $\phi$ and the corresponding four-form $\psi$.
\begin{align}
&d\phi=4\tau_{1}\psi + 3\tau_{7}\wedge\phi -\star\tau_{27} \\
&d\psi=4\tau_{7}\wedge\psi-2\star\tau_{14}
\end{align}
 
From these equations, it follows that the $G_{2}$ three-form having vanishing torsion is equivalent to the exterior derivatives of both the $G_{2}$ three-form and four-form vanishing. In other words, the $G_{2}$ three-form needs to be closed and co-closed. \\

Putting this result together with previous results, the following conditions are all equivalent. 
\begin{enumerate}
\item M has holonomy contained in $G_{2}$ and the holonomy group preserves a positive three-form $\phi$.
\item The $G_{2}$-Structure associated with $\phi$ is torsion-free
\item $\nabla\phi=0$ where $\nabla$ is the Levi-Civita connection
\item d$\phi = \textnormal{d}\star\phi=0$ 
\end{enumerate}

From all the discussion about $G_{2}$-Holonomy, it is clear by now that the $G_{2}$ three-form plays a central role. The correspondence between positive three-forms and $G_{2}$-Structures is also extremely useful, and for this reason the corresponding manifold is often denoted as (M, $\phi$, g) where it is understood that the three-form determines the metric. The term $G_{2}$-Manifold is commonly used in the literature to refer to such a space with a torsion-free connection. 
\begin{definition} $G_{2}$-Manifold. \textnormal{A \textit{$G_{2}$-Manifold} is specified by (M, $\phi$, g) where M is a seven-dimensional Riemannian Manifold and $\phi$ is a torsion-free $G_{2}$ three-form.}
\end{definition}

Note that sometimes in the literature a $G_{2}$-Manifold is also presumed to be compact, but we make no such assumption here. It is also important to keep in mind that the term $G_{2}$-Manifold refers to a space which admits an oriented G$_{2}$-Structure, which only implies that the holonomy is contained in G$_{2}$. In the M-Theory context we will be interested in spaces with holonomy exactly G$_{2}$, so we will need to find additional criteria that ensures we achieve the appropriate holonomy. We will also use the term G$_{2}$-Space to refer to a realistic M-Theory compactification space which is necessarily singular. 

\subsection{Calibrated Geometry}

One challenge regarding the study of $G_{2}$-Manifolds is that there is rarely explicit metric knowledge that can be used for mathematical and physical analysis of these spaces. This lack of explicit metric knowledge is also present in the Calabi-Yau case; however, Yau's proof of the Calabi Conjecture at least ensures existence of such metrics under certain conditions in the compact case \cite{Yau2}. Due to the obvious lack of complex structure in the $G_{2}$ case, there is no known analogous result to the Calabi Conjecture that can be used for existence proofs.\\

Despite the lack of explicit metric knowledge, we can still calculate volumes of certain cycles using the techniques of calibrated geometry developed by Harvey and Lawson \cite{hl}. Being able to calculate these volumes is useful both for the study of M-Theory compactifications and for considerations within differential geometry. In fact, there happens to be a general relationship between the theory of holonomy groups and calibrated geometry. The starting point for introducing calibrated geometry is the notion of an oriented tangent k-plane. 
\begin{definition} Oriented Tangent k-Plane. \textnormal{Let (M, g) be an n-dimensional Riemannian Manifold. An \textit{Oriented Tangent k-Plane} is a vector subspace of dimension $k\leq n$ of T$_{x}$M for $x\in M$ that is equipped with an orientation.}
\end{definition}

Based on the fact that (M, g) is a Riemannian Manifold, the oriented Tangent k-Plane inherits a Euclidean metric from the Riemannian Metric g restricted to it. Since the k-Plane is oriented, the tangent k-Plane V is associated with a natural volume form vol$_{V}$ given in terms of the metric. The notion of a calibration is directly related to this natural volume element. 
\begin{definition} Calibration. \textnormal{A \textit{Calibration} $\phi$ is a closed k-form that satisfies $\left. \phi \right |_{V} \leq \vol_{V}$ for all oriented tangent k-Planes V.}
\end{definition}

The connection between G$_{2}$-holonomy and calibrated geometry follows from the fact that both the $G_{2}$ three-form and its Hodge dual the G$_{2}$ four-form are calibrations. Calibrations themselves are associated with what are called calibrated submanifolds. 
\begin{definition} Calibrated Submanifold. \textnormal{Let N be an oriented k-dimensional submanifold of a Riemannian n-manifold (M, g) and let $\phi$ be a calibration of M. Then each T$_{x}$N is a k-dimensional oriented tangent k-Plane relative to T$_{x}$M. N is a \textit{calibrated submanifold} if $ \left. \phi \right |_{T_{x}N} = \vol_{T_{x}N}$ for all $x\in N$.} 
\end{definition}

When M is a G$_{2}$-Manifold these submanifolds are called associative submanifolds and coassociative submanifolds, corresponding to the G$_{2}$ three-form and G$_{2}$ four-form, respectively. For this reason, the G$_{2}$ three-form is often referred to as the associative three-form and the four-form as the coassociative four-form. More generally speaking, these calibrated submanifolds are significant because they are volume-minimizing within their homology class. This is true globally in the compact case and locally in the non-compact case \cite{JoyceB}. This means that if N and N$'$ are in the same homology class, the volumes satisfy the following relation.  
\begin{equation}
Vol(N)=\int_{N} \phi=\int_{N'} \phi \leq Vol(N')
\end{equation}

As calibrated submanifolds satisfy this condition, they are a subclass of minimal submanifolds. Minimal submanifolds are interesting in and of themselves in the context of differential geometry \cite{cm}, so calibrated geometry certainly has applications in that broader context. In the context of M-Theory compactifications, calibrated geometry is important as a tool for computing particle masses of charged particle states and in conjunction with topological defects such as instantons, domain walls, and strings \cite{hm1}. It is also relevant to keep in mind that the calibration inequality is the BPS inequality in the context of branes wrapped on cycles. \\

\section{Compact Examples}

The Berger Classification constrains the holonomy group of a seven-manifold satisfying the appropriate conditions to be $G_{2}$, but it says nothing about whether or not seven-manifolds with holonomy $G_{2}$ actually exist. Fortunately they do; in fact, every group on Berger's classification is realized as the holonomy group of some appropriate manifold, although Spin(9) is necessarily symmetric \cite{Alexeevy}. Bryant proved the local existence of metrics with $G_{2}$-Holonomy \cite{Bryant} and shortly after Bryant and Salamon explicitly constructed complete non-compact metrics with $G_{2}$-Holonomy \cite{bs}. We will return to the constructions of Bryant and Salamon in the following section, instead first turning our attention to the compact case. As the internal space is oftentimes assumed to be compact, the class of compact $G_{2}$-Manifolds is of particular interest for M-Theory. 

\subsection{Topological Results}

Compact manifolds of holonomy $G_{2}$ satisfy many topological properties that do not hold generally. These are often useful for the construction of these manifolds and are extremely important when we study M-Theory compactified on these spaces. One preliminary result is that compact seven-manifolds admitting metrics of holonomy $G_{2}$ are automatically orientable and spin \cite{JoyceB}. It is spin as a direct consequence of the fact that G$_{2}$ $\subset$ SO(7) is simply-connected. Another particularly important result concerns when the holonomy group of a compact manifold with a torsion-free G$_{2}$-Structure is exactly G$_{2}$. From the point of view of constructing G$_{2}$-Manifolds starting from considerations of G$_{2}$-Structure, identifying a torsion-free G$_{2}$-Structure is only equivalent to the holonomy of the manifold being contained in G$_{2}$. A necessary and sufficient criterion for the holonomy being exactly G$_{2}$ is given in terms of the fundamental group of the manifold. 
\begin{thm} Holonomy Group and Fundamental Group. \textnormal{Let (M, g) be a compact Riemannian Manifold with holonomy contained in $G_{2}$. The holonomy group of the manifold is exactly $G_{2}$ if and only if the fundamental group of M is finite.}
\end{thm} 

This result is used directly in every construction of compact G$_{2}$-Manifolds, and there is a distinct result that is applicable to the non-compact case for exactly determining the holonomy. In the compact case we can also decompose the de Rham Cohomology Groups in terms of refined groups whose dimensions correspond to refined Betti numbers. 
\begin{definition} Refined Betti Numbers. \textnormal{Let $\Lambda_{l}^{k}$(M) correspond to the irreducible components of $\Lambda^{k}$(M) as defined in equations 1.9-1.14. Let H$_{l}^{k}$ denote the vector subspace of $\Lambda_{l}^{k}$(M) consisting of closed and co-closed k-forms. The \textit{Refined Betti Number} b$_{l}^{k}$ is the dimension of the de Rham Cohomology group associated to the vector subspace H$_{l}^{k}$. }
\end{definition}

We can then decompose the de Rham Cohomology Groups in terms of the groups associated with the refined Betti numbers in the following way. 
\begin{thm} de Rham Decomposition. \textnormal{Let (M, $\phi$, g) be a compact seven-manifold with holonomy contained in G$_{2}$. Then we can decompose the de Rham Cohomology groups in terms of the refined de Rham Cohomology groups described above as presented below for each degree of k-form.} 
\begin{align}
&H^{2}(M,\mathbb{R})=H_{7}^{2}(M, \mathbb{R}) \oplus H_{14}^{2}(M, \mathbb{R})\\
&H^{3}(M,\mathbb{R})=H_{1}^{3}(M, \mathbb{R}) \oplus H_{7}^{3}(M, \mathbb{R}) \oplus H_{27}^{3}(M, \mathbb{R})\\
&H^{4}(M,\mathbb{R})=H_{1}^{4}(M, \mathbb{R}) \oplus H_{7}^{4}(M, \mathbb{R}) \oplus H_{27}^{4}(M, \mathbb{R})\\
&H^{5}(M,\mathbb{R})=H_{7}^{5}(M, \mathbb{R}) \oplus H_{14}^{5}(M, \mathbb{R})
\end{align}

\textnormal{If the holonomy of (M, $\phi$, g) is exactly G$_{2}$ then $H_{7}^{k}$ are trivial for k between 1 and 6.}

\end{thm}

In the compact case we can also prove a statement regarding the first Pontryagin class of M, denoted p$_{1}$(M). The k$^{th}$ Pontryagin class satisfies $\text{p}_{k} (\text{M})\in \text{H}^{4k}$(M, $\mathbb{Z}$) and since we are interested in G$_{2}$-Manifolds which have dimension seven, p$_{1}$(M) will not automatically vanish. In fact, using Chern-Weil Theory we can show that it does not vanish \cite{JoyceB}.  
\begin{thm} Compact $G_{2}$-Manifolds and First Pontryagin Class. \textnormal{Let (M, g) be a compact $G_{2}$-Manifold. Then its first Pontryagin Class is nonzero.}
\end{thm}

In the context of M-Theory Compactifications, p$_{1}$(M) comes up in the context of a cohomological consistency condition for the supergravity action to be consistent at one loop. 

\subsection{Orbifold Resolution Constructions}

The first examples of compact $G_{2}$-Manifolds were constructed by Joyce in 1996 \cite{Joyce1,Joyce2} and later generalized by him in 2000 \cite{JoyceB}. The method he used was analogous to the Kummer Construction for K3 Surfaces. The goal of this construction is to identify a particular K3 Surface and show that it admits a hyperk{\"a}hler structure. Explicit K3 metric knowledge is not known in general, but a particularly advantageous class of spaces to consider are those that correspond to certain limits of the boundary of K3 moduli space \cite{Huybrechts}. The spaces corresponding to such limits are orbifolds of the four-torus.  \\

Based on these considerations, the starting point for the Kummer Construction is the orbifold $T^{4}/~\Gamma$, where $\Gamma=\mathbb{Z}_{2}$ is the finite group corresponding to isometric involutions of the four-torus that reverse directions. Under the action of this finite group, the orbifold has 16 singular points. We can resolve these singularities by patching them with Eguchi-Hanson Spaces. 
\begin{definition} Eguchi-Hanson Space. \textnormal{An \textit{Eguchi-Hanson Space} is a non-compact hyperk{\"a}hler manifold where the space is the cotangent bundle of complex projective space and the metric is complete.}
\end{definition}

Ultimately, the orbifold $T^{4}/~\Gamma$ with the singularities resolved by Eguchi-Hanson Spaces will be the manifold identified by the Kummer Construction that admits a hyperk{\"a}hler structure. However, at the current stage the resolved orbifold exhibits nearly hyperk{\"a}hler structure. It is at this point in the construction that the choice to resolve the orbifold singularities with Eguchi-Hanson Spaces becomes important: the degree to which the resolved orbifold fails to be hyperk{\"a}hler can effectively be tuned via the Eguchi-Hanson metrics \cite{Page}. This fact motivates the use of Eguchi-Hanson Spaces to resolve the singularities, though in principle any space could be used to resolve the singularities provided that the resolutions can be used to tune the structure of the manifold. \\

Given that the K3 surface is nearly hyperk{\"a}hler and tunable via the Eguchi-Hanson metrics, it can be shown that the K3 surface specified in the construction admits hyperk{\"a}hler structure via techniques from deformation theory and twistor theory of singular complex manifolds \cite{ls}. \\

As was the case with the K3 moduli space, certain limits of the boundary of the $G_{2}$ moduli space are particularly amenable to study. These limits once again correspond to orbifolds, but this time of the seven-torus under the action of a different finite group of involutions. The following results specialize to Joyce's original 1996 construction in order to present a more intuitive picture of the construction. The generalization effectively increases the number of singularities that can be resolved, and thereby increases the choices of $\Gamma$ that may be associated with the orbifold singularities. The important feature of the singularities in the general case is that they must be modeled on SU(3) singularities in order to be resolved via these methods. \\ 

In the specific case, the finite group is generated by the maps $\alpha$, $\beta$, and $\gamma$ from $\mathbb{R}^7=\mathbb{R}^3 \oplus \mathbb{C}^2$ to itself specified by their actions on coordinates as specified below. 
\begin{align}
& \alpha(x_{1},x_{2},x_{3},z_{1},z_{2})=(x_{1},x_{2},x_{3},-z_{1},-z_{2}) \\
& \beta(x_{1},x_{2},x_{3},z_{1},z_{2})=(x_{1},-x_{2},-x_{3},iz_{2},-iz_{1}) \\
& \gamma(x_{1},x_{2},x_{3},z_{1},z_{2})=(-x_{1},x_{2},-x_{3},\overline{z_{1}},\overline{z_{2}}) 
\end{align}

 Note that this group of involutions is precisely the collection of transformations that preserve the flat $G_{2}$-Structure on the seven-torus. Joyce's motivation for generalizing the Kummer Construction followed from these parallels of the limits of the boundary of the moduli spaces. The basic steps of the Kummer Construction in the K3 and $G_{2}$ cases are similar: Starting from the aptly-chosen orbifold, we resolve the singularities with Eguchi-Hanson Metrics and can show that the resulting manifold has a nearly torsion-free $G_{2}$-Structure, which is tunable via the Eguchi-Hanson metrics. However, the Kummer Construction for K3 surfaces used results from the theory of singular complex manifolds to ultimately prove that the spaces admit hyperk{\"a}hler structure, but in the $G_{2}$ case there is no complex structure to work with so the deformation theory has to be built up from scratch and predominately relies on techniques from analysis. Much of Joyce's work towards generalizing the Kummer Construction consisted of developing the deformation theory for $G_{2}$-Structures, which ultimately is very useful in and of itself as well as applied to other constructions of $G_{2}$-Manifolds. Regarding the compactness, note that the space $T^{7}/~\Gamma$ is compact following from the compactness of the torus and the fact that the action of the group, the subsequent resolving of singularities, and the deformation theory ultimately preserve the compactness of the overall manifold. \\
 
 Given these considerations, we are able to state Joyce's existence theorem for G$_{2}$-Metrics. In addition to the specification of the orbifold $T^7/~\Gamma$ the existence result requires the specification of Resolution-Data (R-Data). Roughly speaking, this R-Data is the information needed to associate a resolution M with the orbifold \cite{JoyceB}. In terms of R-Data then, the existence result is as follows. 
\begin{thm} Joyce Construction. \textnormal{Let T$^{7}/~\Gamma$ have a flat G$_{2}$-Structure and let it have a set of R-Data. In addition, let M be the corresponding resolution of the orbifold and let it have finite fundamental group. Then M admits metrics with holonomy G$_{2}$.}
\end{thm}

Now that we have this existence result for G$_{2}$-Metrics on specific resolutions of orbifolds, we can investigate topological properties of the resolutions. In particular we are interested in the most fundamental topological invariants, namely the fundamental group and the Betti numbers. By the assumption of the theorem the fundamental group is finite, so as a consequence b$^{1}(\text{M})=0$. Moreover, since the manifolds of the Joyce construction are compact and orientable, the Betti numbers are related in the usual fashion so we only need to compute b$^{2}$(M) and b$^{3}$(M). The third Betti number is especially significant based on its relationship with the space of G$_{2}$ three-forms, but both of these Betti numbers are significant in the context of studying supergravity on Joyce manifolds. \\

We can identify simply-connected examples of the Joyce construction by requiring that the fundamental group not only be finite but also be trivial. Based on the eight examples of Joyce orbifold resolutions corresponding to Joyce's generalized construction, there are 252 sets of second and third Betti numbers corresponding to simply-connected Joyce manifolds and therefore at least that many distinct compact G$_{2}$-Manifolds \cite{JoyceB}. For the constructed spaces b$^{2}(M)$ ranges between 0 and 28 and b$^{3}(M)$ between 4 and 215. \\

More recently, Joyce and Karigiannis have developed a distinct construction that also works by resolving orbifold singularities \cite{jk}. Instead of starting with the seven-torus and acting on it with a finite group, they assume from the onset that a Riemannian Manifold (M, g) admits an involution i which preserves the G$_{2}$-Structure. Then $\text{M}/\langle i \rangle$ is a G$_{2}$-Orbifold which can be resolved via Eguchi-Hanson spaces and deformed to a smooth compact G$_{2}$-Manifold analogously to the procedure in Joyce's original construction. Compactness of these spaces is assumed from the onset as opposed to being a consequence of starting with the seven-torus.

\subsection{Twisted Connected Sum Construction}

Kovalev, following a suggestion of Donaldson, developed the next construction of compact G$_{2}$-Manifolds in 2003 \cite{Kovalev}, and these spaces (and their generalizations) along with those of Joyce and Joyce/Karigiannis comprise our current examples of compact G$_{2}$-Manifolds. Kovalev's construction may be thought of as a generalization of a connected sum construction \cite{connectedsum} designed to ensure that the fundamental group of the resulting manifold is finite, thereby ensuring that the holonomy is exactly G$_{2}$. He was able to achieve this constraint on the fundamental group by ``twisting'' the building blocks used in the generic connected sum, and for this reason this method  is commonly called the twisted connected sum construction. \\

Kovalev used spaces of holonomy SU(3) in his construction because they are relatively well-understood from their context in algebraic geometry. Tian and Yau had already found existence theorems for Ricci-Flat K{\"a}hler manifolds on quasiprojective spaces \cite{ty}, but this construction requires additional structure on the spaces to ensure that the gluing procedure functions appropriately. To this end, Kovalev began his construction by generalizing these results to prove the existence of Ricci-Flat K{\"a}hler metrics with holonomy SU(3) that were also asymptotically cylindrical. 
\begin{definition} Calabi-Yau Cylinder. \textnormal{Let V= $\mathbb{C^{*}}$$\times$ S where S is a K3 surface associated with Ricci-Flat K{\"a}hler form $\omega$ and holomorphic two-form $\Omega$. Choose a complex coordinate z = $e^{t+i\theta}$. Then specify a Ricci-Flat K{\"a}hler form $\omega '$ and a holomorphic two-form $\Omega '$ on V via the following equations.}

\begin{align}
&\omega'=dt \wedge d\theta + \omega \\
&\Omega'=(d\theta-i dt) \wedge \Omega
\end{align}

\textnormal{Then (V, $\omega', \Omega'$) is defined to be a \textit{Calabi-Yau Cylinder} and it comes equipped with a projection $\xi : \text{V} \rightarrow \mathbb{R}$ given by $\xi (\text{z},\text{x})=\log\abs{z}$ }
\end{definition}

This notion of a Calabi-Yau Cylinder features in the definition of an Asymptotically Cylindrical Calabi-Yau three-fold, namely, the asymptotically cylindrical Calabi-Yau three-fold approaches a Calabi-Yau Cylinder in an asymptotic limit. The gluing associated with the twisted connected sum construction occurs in this asymptotic limit. 
\begin{definition} Asymptotically Cylindrical. \textnormal{Let X be a complete Calabi-Yau three-fold with Ricci-Flat K{\"a}hler form $\omega$ and holomorphic three-form $\Omega$. V is called an \textit{Asymptotically Cylindrical Calabi-Yau three-fold} if there is a compact set Y $\subset$ X, a Calabi-Yau Cylinder (V, $\omega' $, $\Omega'$) with associated projection $\xi$, and a diffeomorphism $\eta$ : $\xi^{-1}$(0, $\infty$) $\rightarrow$ V $\backslash$  \textnormal{Y that satisfies}} \\

\begin{align}
&\eta^{*}\omega-\omega'=d\rho \\
&\eta^{*}\Omega-\Omega'=d\zeta
\end{align}

\textnormal{In these expressions, the forms $\rho$ and $\zeta$ are presumed to satisfy the following boundedness properties given in terms of the derivative operator and norms associated with the Calabi-Yau metric on the Calabi-Yau cylinder for all k $\geq$ 0, for some $\lambda$, and as t $\rightarrow \infty$}

\begin{align}
&\mid \nabla^{k}\rho \mid = O(e^{\lambda t}) \\
&\mid \nabla^{k}\zeta \mid = O(e^{\lambda t}) 
\end{align}

\end{definition}

Note that in this definition it is the parameter t which characterizes the asymptotic limit, and by construction, the Asymptotically Cylindrical Calabi-Yau approaches the Calabi-Yau Cylinder in this limit involving t. As stated previously, Kovalev's first result is an existence theorem for these asymptotically cylindrical Calabi-Yau three-folds. 
\begin{thm} Kovalev Construction of Asymptotically Cylindrical Calabi-Yau Three-Folds. \textnormal{Let M be a smooth compact K{\"a}hler three-fold with H$^{1}$(M, $\mathbb{R})=0$. Let X be a K3 surface that is an anticanonical divisor of M with trivial self-intersection class N $\cdot$ N = 0 in $H_{2}(M,\mathbb{Z})$. Then the space M$\slash$X admits an asymptotically cylindrical Ricci-Flat K{\"a}hler Metric. If in addition M is simply-connected and has a complex curve $\ell$ that satisfies X $\cdot$ $\ell>$ 0 then the asymptotically cylindrical Ricci-Flat K{\"a}hler metric  has holonomy SU(3).}
\end{thm}

These Asymptotically-Cylindrical Calabi-Yau three-folds are non-compact and have six real dimensions. They are related to Fano three-folds, which are well-known in the context of algebraic geometry. We present the definition below for readers with some background in the discipline. 
\begin{definition} Fano Variety. \textnormal{A \textit{Fano Variety} is a complete variety whose anticanonical class is ample.}
\end{definition}

The fact that Kovalev uses Asymptotically Cylindrical Calabi-Yau three-folds derived from Fano three-folds will be important when we discuss generalizations to the Twisted Connected Sum construction. To construct a seven-dimensional space, we may take the product of one of these three-folds with a circle. Doing so preserves the SU(3) holonomy and as SU(3) is a subgroup of $G_{2}$ these spaces have holonomy contained in $G_{2}$. Asymptotically, these seven-dimensional spaces become the product of the circle with the Calabi-Yau Cylinder associated to the three-fold. In this asymptotic limit, the cross-section of the cylinder is of the form $S_{1}$ $\times$ $S_{1}$ $\times$ K where K is a K3 surface with hyperk{\"a}hler structure. \\

These spaces built up from Fano three-folds are the building blocks for the twisted connected sum construction. The next step is to glue together any two such seven-manifolds that obey an appropriate matching condition, called the Donaldson Matching Condition. 
\begin{definition} Donaldson Matching Condition. \textnormal{Two seven-dimensional building blocks A and B with neck cross sections (K$_{A}$, S$_{A^1}$, S$'_{A^1}$) and (K$_{B}$, S$_{B^1}$, S$'_{B^1}$) are said to satisfy the \textit{Donaldson Matching Condition} if there is a diffeomorphism r between the K3 Surfaces K$_{A}$ and K$_{B}$ that preserves the Ricci-Flat metric and relates the hyperk{\"a}hler structures ($\omega$, $\Omega$) in the following ways: }

\begin{align}
& r^{*}(\omega_{B})=\Re(\Omega_{A}) \\
& r^{*}(\Re(\Omega_{B}))=\omega_{A} \\
& r^{*}(\Im(\Omega_{B}))=-\Im(\Omega_{A})
\end{align}

\end{definition}

In Kovalev's original paper he defines a matching condition between K3 surfaces as opposed to defining a matching condition resembling Donaldson Matching. However, the matching conditions he places on the K3 surfaces imply the existence of the diffeomorphism known as Donaldson Matching as a consequence of the global Torelli Theorem for K{\"a}hler K3 Surfaces \cite{Huybrechts}, so the end results are the same. Additionally, the term hyperk{\"a}hler rotation is commonly used to denote the Donaldson Matching procedure; however, Donaldson Matching properly speaking is a precise case of a hyperk{\"a}hler rotation \cite{hm1}. \\

Provided that the K3 surfaces associated with the building blocks satisfy the Donaldson Matching Condition, there exists another diffeomorphism that can be used to effectively glue the seven-manifolds together in their asymptotic regions in such a way as to produce a compact seven-manifold with an approximately torsion-free $G_{2}$-Structure. The procedure for gluing involves truncating the ends and identifying the boundaries  $S_{1} \times S_{1} \times$ K via an orientation reversing isometry. In terms of the identification this means that the circles associated with the boundaries of the two surfaces are interchanged and the complex structures of the spaces K are interchanged via the aforementioned global Torrelli Theorem. \\

The reason for gluing together the boundaries in this twisted sense is to ensure that the holonomy of the resulting spaces is exactly $G_{2}$. The twisting procedure achieves this by ensuring that the fundamental group of the manifold is finite. If the circles were mapped into each other without the twist the fundamental group would be infinite corresponding to the fundamental group for the circle being $\mathbb{Z}$. \\

The compact seven-manifolds produced via this method have an almost torsion-free $G_{2}$-Structure which can be made exactly torsion free using the deformation theory developed by Joyce. Noting both the torsion-free $G_{2}$-Structure and finite fundamental group, it follows that the manifolds produced have holonomy $G_{2}$. To give perspective to the number of manifolds produced by this method, there are hundreds of Asymptotically Cylindrical Calabi-Yau three-folds associated to Fano three-folds. Then we would expect for there to be hundreds of corresponding compact G$_{2}$-Manifolds. \\

Topologically, the features of the twisted connected sum spaces are rather distinct from those of Joyce's orbifold resolutions. For instance, the Betti numbers are related, but not uniquely determined by, those of the building blocks \cite{Kovalev}. Oftentimes, the second Betti number vanishes, but it is necessarily true that 0 $\leq$ b$_{2}$(M)$ \leq$ 9. If M, V$_{1}$, and V$_{2}$ label the G$_{2}$-Manifold and the two summands respectively, then the third Betti number of the G$_{2}$-Manifold satisfies 
\begin{equation}
b_{3}(M)+b_{2}(M)=b_{3}(V_{1})+b_{3}(V_{2})+23
\end{equation}

For the examples of twisted connected sums given in Kovalev's paper, he realizes 71 $\leq$ b$_{3}$(M) $\leq$ 155. \\

Implicit in Kovalev's association of Fano three-folds to building blocks in the twisted connected sum construction is the possibility for generalization to develop new examples of compact manifolds of holonomy $G_{2}$. In practice both the choice of building blocks for the twisted connected sum construction and the matching of these building blocks can be generalized. One approach for generalization involves using as building blocks K3 surfaces with non-symplectic involutions \cite{kl}. Another involves using weak Fano three-folds as opposed to Fano three-folds for the building blocks \cite{chnp1}. 
\begin{definition} Weak Fano-Variety. \textnormal{A \textit{Weak Fano Variety} is a complete variety whose anticanonical class is big and nef, but not ample.}
\end{definition}

Corti-Haskins-Nordstr{\"o}m-Pacini first develop an existence proof for Asymptotically Cylindrical Calabi-Yau three-folds from Weak Fano three-folds \cite{chnp2}. Using these Weak Fano three-folds as opposed to Fano three-folds increases the number of building blocks from several hundred to several hundred thousand. \\

Despite being interesting for the sake of constructing asymptotically cylindrical Calabi-Yau three-folds, not all of these weak Fano three-folds are suitable building blocks for the twisted connected sum to produce compact $G_{2}$-Manifolds. Roughly speaking, the reason for this incompatibility is that they do not have the appropriate deformation theory, or at least, they are not appropriate for the deformation theory already developed by Joyce. So for the sake of using these new spaces in a twisted connected sum, identifying a subclass of weak Fano three-folds that do have the correct deformation theory is important. This subclass consists of the Semi-Fano three-folds. The precise definition is rather technical, but the main idea is that they satisfy certain cohomological conditions that are conducive towards using the deformation theory for G$_{2}$-Structures. \\

Following this generalization to building blocks derived from Semi-Fano three-folds, it is much more difficult to determine the exact possibilities for G$_{2}$-Manifolds, as Semi-Fano three-folds, or even weak Fano three-folds, are not classified to the extent that Fano three-folds are \cite{chnp1}. Even so, considering just a subset of possible two-connected G$_{2}$-Manifolds, they are able to realize a wide range of values for the third Betti number, precisely 55 $\leq$ b$_{3}$(M) $\leq$ 239. These two-connected examples have a strict lower bound of 31 for the third Betti Number. \\

More importantly however, they developed techniques to compute the full integral cohomology for G$_{2}$-Manifolds of twisted connected sum type, which is particularly helpful for distinguishing G$_{2}$-Manifolds based on topological invariants. In particular, they demonstrate that the same building blocks may lead to topologically-distinct G$_{2}$-Manifolds, and they are able to classify homeomorphism and diffeomorphism types for certain spaces of the twisted connected sum type. 

\section{Non-Compact Examples}

While oftentimes physicists assume that the M-Theory compactification space is compact, non-compact examples are still important to physics in two respects. On the one hand, non-compact complete $G_{2}$-Manifolds are in principle still viable possibilities for compactification spaces so long as the Laplacian remains well-behaved (with regards to spectral theory considerations) on these spaces. One advantageous feature of compact spaces is that they ensure that the Laplacian is well-behaved in this respect; however, complete non-compact spaces in principle could still be well-behaved. Non-compact examples are also extremely important because in the context of physically-realistic M-Theory compactifications, it will be necessary to introduce singularities into the internal space. Lacking examples of compact spaces with appropriate singularities, physicists study M-Theory locally near these singularities in order to understand the fundamental aspects of the physics. These local models are non-compact G$_{2}$-Manifolds which have specific types of degenerations.  

\subsection{Considerations in the Non-Compact Case}

It is important to note that many topological results that held in the compact case, for instance those discussed in the section on compact G$_{2}$-Manifolds, no longer apply. In particular, we can no longer use the fundamental group argument in relation to the holonomy of the manifold, which was instrumental in the constructions of Joyce and Kovalev. However, there is a distinct criterion that we can use to ensure that the holonomy of the space is exactly G$_{2}$, provided the manifold satisfies certain topological conditions \cite{bs}. 
\begin{thm} Simply-Connected Manifolds and Holonomy G$_{2}$. \textnormal{Let (M, g) be a Riemannian Manifold where M is simply-connected and Hol(g) $\subseteq$ G$_{2}$. Then $\Hol(g)=G_{2}$ if and only if there are no non-trivial parallel one-forms under the Levi-Civita connection.}
\end{thm}

Note that this result applies in the (simply-connected) compact case as well but it is easiest to deal with the fundamental group in that context. However, there are technical advantages for constructing non-compact examples that are not shared by the compact case. For instance, in the compact case there are no continuous symmetries that may be used in constructions. This fact is a consequence of the following theorem. 
\begin{thm} Compact G$_{2}$-Manifolds and One-Forms. \textnormal{Let (M, g) be a Compact Ricci-Flat Riemannian Manifold and let $\xi$ be a one-form. Then $\nabla\xi=0$. In addition, all one-forms are constant and the holonomy is reducible.}
\end{thm}

As a direct result of this theorem, compact, irreducible G$_{2}$-Manifolds (which are automatically Ricci-Flat) do not admit any non-trivial one-forms whose duals could be associated with Killing vectors. Therefore, there can be no continuous symmetries in such cases. However, in the non-compact case we can get around this result and may expect to be able to use continuous symmetries. In fact, basically all complete non-compact G$_{2}$-Manifolds constructed thus far admit a Lie Group action with generic orbit of codimension one \cite{fhn1}. Spaces with these types of Lie Group actions are called cohomogeneity one spaces. \\

\subsection{Specific Constructions}

The first examples of non-compact complete $G_{2}$-Manifolds were constructed by Bryant and Salamon \cite{bs}. They were able to explicitly construct complete G$_{2}$-Holonomy metrics on three specifically-chosen non-compact manifolds.  All of the G$_{2}$-Holonomy manifolds they constructed were total spaces of vector bundles of rank three or four. The one example from the rank three case was defined on the spin bundle over $S^{3}$ and the two other examples of rank four were defined on $\Lambda_{-}^{2}(M)$ where M is either $S^{4}$ (with standard metric) or $\mathbb{C}P^{2}$ (with Fubini-Study metric) and $\Lambda_{-}^{2}(M)$ denotes the space of anti-dual two-forms on M. An important feature to note about all of these examples is that they are asymptotically conical. Such spaces are defined in relation to Calabi-Yau Cones \cite{fhn1}.
\begin{definition} Calabi-Yau Cone. \textnormal{Let (M, g$_{M}$) be a Riemannian Manifold and let C(M) denote the incomplete space (0, $\infty$) $\times$ M with a metric of the following form.} \\

\begin{equation}
g_{C(M)}=dt^2 + t^{2} g_{M}
\end{equation}

\textnormal{C(M) is called a \textit{Calabi-Yau Cone} if the metric g$_{C(M)}$ is induced via a Calabi-Yau structure on C(M).}
\end{definition}

In this definition, the Riemannian Manifold M (typically compact) is called the base of the cone. Note that via the metric ansatz the Calabi-Yau Cone is a singular space. Asymptotically Conical spaces, as their name suggests, are smooth spaces related to these singular spaces in an asymptotic sense. 
\begin{definition} Asymptotically Conical (AC). \textnormal{Let X be a complete Calabi-Yau three-fold and let Y be a compact subset of X. Then X is an \textit{Asymptotically Conical Calabi-Yau Three-Fold} if X $\slash$ Y is diffeomorphic to a three-dimensional Calabi-Yau Cone.}
\end{definition}

Following from the definitions, these AC Spaces may be thought of as resolutions of conical singularities, which is why they are so useful for modeling local physics. These types of isolated singularities will come up in the physical context of chiral fermions later. \\

The next examples of cohomogeneity-one complete non-compact G$_{2}$-Manifolds came from the physics community. The ability to use continuous symmetries in the non-compact case is very conducive for use in physics since one could build up spaces with symmetry groups appropriate for use in specific physical theories. These approaches use the dualities of M-Theory with other superstring theories to motivate specific choices of symmetry groups. As we will see throughout the review, such dualities are often very useful for probing physical features of M-Theory. The physical considerations surrounding the first construction involved the M-Theory lift of N D6-Branes wrapping the S$^3$ in the deformed conifold geometry of type IIA string theory \cite{bggg}. The symmetry group corresponding to this physical setup in the superstring theory may be extrapolated from the equation for a deformed conifold. 
\begin{equation}
x_{1}^2+x_{2}^2+x_{3}^2+x_{4}^2=r
\end{equation}

This equation has a SU(2) $\times$ SU(2) symmetry associated with rotating the arguments and in the case that r does not vanish, it also has a $\mathbb{Z}_{2}$ symmetry associated with changing the signs of the variables. Noting then that the M-Theory lift is further associated with a U(1) symmetry corresponding to the M-Theory circle, the symmetry imposed on the compactification space should be SU(2) $\times$  SU(2) $\times$  U(1) $\times$ $\mathbb{Z}_{2}$. \\

The most general metric invariant under these symmetries has vielbeins that depend on four functions of a fixed radial coordinate.  Since the $G_{2}$ three-form and its Hodge dual can be expressed locally in the vielbein basis, we can directly ensure that the holonomy is contained in G$_{2}$ by requiring that the G$_{2}$ three-form be closed and co-closed. Imposing these conditions results in a system of coupled first-order differential equations for the four functions (denoted A, B, C, and D) appearing in the vielbeins. 
\begin{equation}
\frac{dA}{dr}=\frac{1}{4}\left(\frac{B^2-A^2+D^2}{BCD}+\frac{1}{A}\right)
\end{equation}
\begin{equation}
\frac{dB}{dr}=\frac{1}{4}\left(\frac{A^2-B^2+D^2}{ACD}+\frac{1}{B}\right)
\end{equation}
\begin{equation}
\frac{dC}{dr}=\frac{1}{4}\left(\frac{C}{B^2}-\frac{C}{A^2}\right)
\end{equation}
\begin{equation}
\frac{dA}{dr}=\frac{1}{2}\frac{A^2+A^2-D^2}{ABC}
\end{equation}

Alternatively, these equations may be realized by starting with the Ricci-Flatness condition and then utilizing the aforementioned symmetry group. It is important to note that these equations are meant to determine a metric without reference to a specific space on which to put it. The idea in using this approach to construct complete non-compact G$_{2}$-Manifolds is that if we put such a metric on an appropriate simply-connected space then there is a straightforward criterion for determining its holonomy using a theorem from the previous section. Namely, since the holonomy is contained in G$_{2}$, we can show the holonomy is G$_{2}$ by proving there are no non-trivial parallel one-forms. \\

Having clarified this point, we can consider these equations specifically. They are noteworthy because not only do they reproduce the metrics of Bryant and Salamon in an appropriate limit, but they also suggest the existence of an entirely new one-parameter family of cohomogeniety-one G$_{2}$-Metrics. \\

This construction is actually just one of four one-parameter families of cohomogeniety-one G$_{2}$-Metrics. One aspect in which they differ is in regards to their symmetry groups. The family discussed previously is denoted $\mathbb{B}_{7}$, and the other families are $\mathbb{A}_{7}$ \cite{a7}, $\mathbb{C}_{7}$ \cite{cglp1}, and $\mathbb{D}_{7}$ \cite{cglp2}. Up to discrete symmetries, the families $\mathbb{C}_{7}$ and $\mathbb{D}_{7}$ also have symmetry group SU(2) $\times$  SU(2) $\times$  U(1) and the family $\mathbb{A}_{7}$ has symmetry group SU(2) $\times$  SU(2) \cite{fhn1}. \\

These families are also related via their asymptotic geometry. Whereas the Bryant-Salamon constructions were asymptotically cylindrical, these four families of codimension-one G$_{2}$-Metrics are called Asymptotically Locally Conical (ALC). Roughly speaking, ALC spaces asymptotically resemble the product of a twisted circle and a six-dimensional AC space.\\
 
 Analogously to the approach for the $\mathbb{B}_{7}$ case, systems of ordinary differential equations provide the informal justification for these families of solutions. The existence of solutions for these ordinary differential equations systems have been proven rigorously for the families $\mathbb{B}_{7}$ \cite{B,fhn1}, $\mathbb{C}_{7}$\cite{fhn1}, and $\mathbb{D}_{7}$\cite{fhn1}. In addition to proving the existence of the families $\mathbb{C}_{7}$ and $\mathbb{D}_{7}$ for the first time, these authors also constructed infinitely many new one-parameter families of complete simply-connected cohomogeneity-one G$_{2}$-Manifolds all with symmetry group SU(2) $\times$  SU(2) $\times$  U(1) \cite{fhn1}. \\ 
 
As noted in the $\mathbb{B}_7$ case, each of these families has a limit in which the ALC geometry at infinity corresponds to one of the AC Bryant-Salamon examples. Crucially though, they have another limit that corresponds to the collapse of the ALC G$_{2}$-Metric to a metric on a AC Calabi-Yau three-fold. This limit affords a possibility to construct additional examples of complete non-compact $G_{2}$-Metrics by effectively reversing this limit, starting with an AC Calabi-Yau and using it (with additional data) to construct an ALC $G_{2}$-Manifold \cite{fhn2}. The practical advantage of this approach follows from the fact that there are many examples of AC Calabi-Yau metrics so presumably such a procedure would produce many complete non-compact ALC G$_{2}$-Metrics. A significant distinction of this approach from those concerning the established codimension-one families is that in the latter case the high degree of symmetry allowed for the equations determining G$_{2}$-Metrics to reduce to ordinary differential equations. In the case of this construction, such symmetry is not anticipated so the equations are partial differential equations. \\

In this case the PDE system corresponds to what are called the Apostolov-Salamon equations and the seven-manifold corresponding to the complete, non-compact, and simply-connected G$_{2}$-Manifold is a U(1)-bundle of a simply-connected AC Calabi-Yau three-fold over a five-dimensional Sasaki-Einstein Manifold (with additional topological constraints). They demonstrate existence of G$_{2}$-Holonomy metrics on these spaces by solving the linearized Apostolov-Salamon equations by considering a power series solution. Ultimately, they are able to prove the existence of continuous families of G$_{2}$-Manifolds with arbitrarily high numbers of parameters, complementing the one-parameter families originally motivated by physicists. \\

Using the AC to ALC limit is particularly useful because many AC Calabi-Yau threefolds are already known. However, these spaces are not adequately categorized so it is difficult to specify exactly how many G$_{2}$-Manifolds follow from this construction \cite{fhn2}.  

\section{$G_{2}$ Moduli Space}

In addition to studying specific constructions of G$_{2}$-Manifolds, another direction that is interesting to take regarding $G_{2}$-Holonomy is to study the moduli space of G$_{2}$-Metrics on a given seven-manifold. This moduli space is an important construct in its own right and characteristics of the moduli space could potentially be used to prove results relating to specific $G_{2}$-Manifolds. 

\subsection{Local Aspects} 

The moduli space we will spend the most time discussing is that of G$_{2}$-Structures on a given smooth compact manifold with fixed topology. Keeping in mind the relationship between G$_{2}$-Structures and positive three-forms, we can characterize the moduli space directly in terms of the space of smooth positive three-forms, denoted $P^3(M)$. When we define the moduli space we will want these $G_{2}$-Structures to be torsion-free and for the points in the moduli space to be defined up to diffeomorphism. We capture both of these notions in the following definition. 
\begin{definition} Moduli Space of Torsion-Free $G_{2}$-Structures. \textnormal{Let (M, g) be a compact, oriented seven-manifold. Define a set X to be the space of smooth positive three-forms, denoted $P^3(M)$, corresponding to torsion-free $G_{2}$-Structures}.

\begin{equation}
X=\{\phi \in P^3(M) : d\phi=d\star\phi=0\}
\end{equation}

\textnormal{Define D to be the group of diffeomorphisms isotopic to the identity, for which there is a natural action of D on X defined via the pullback. The \textit{Moduli Space of torsion-free $G_{2}$-Structures} then corresponds to the quotient X/D.}
\end{definition}

The mathematical structure of the torsion-free moduli space is not immediately apparent from the definition. However, Joyce proved that it is possible to identify a ``slice'' of the action of D on X with a submanifold of X \cite{JoyceB}. The existence of this slice has strong implications for the local nature of the $G_{2}$ Moduli Space. 
\begin{thm} Local Structure of $G_{2}$ Moduli Space. \textnormal{Let (M, g) be a compact manifold and let M be the moduli space of torsion-free $G_{2}$-Structures. Then M is a smooth manifold of dimension $b^{3}(M)$ and M is locally diffeomorphic to H$^3$(M, $\mathbb{R}$).}
\end{thm}

We can also arrive at this result through what is called the Hitchin Functional \cite{Hitchin1}. This approach has major ramifications for the theory not only for exceptional holonomy, but also in the Calabi-Yau case. In the context of $G_{2}$-Holonomy, we are interested in stable three-forms \cite{Hitchin2}.
\begin{definition} Stable. \textnormal{A k-form is called \textit{stable} if the orbit of this form under the action of the general linear group on $\Lambda^{k}(M)$ is open.}
\end{definition}

Stability is important in this context because it will allow us to define a functional whose variation is well-defined. Let this functional be expressed in the following form.
\begin{equation}
F(\Phi)=\int_{M} \Phi \wedge \psi
\end{equation}

Here, $\psi$ is the Hodge star of $\Phi$. If we perform a variation, which is well-defined since the forms are stable, we recover the condition that the three-form be closed and co-closed. As we saw earlier, this condition is equivalent to the holonomy being contained in G$_{2}$. Amazingly, the converse is also true if we restrict the functional to the cohomology class of $\Phi$. 
\begin{thm} Hitchin Functional and G$_{2}$-Holonomy. \textnormal{Let M be a Riemannian seven-manifold with holonomy G$_{2}$ associated to G$_{2}$ three-form $\Phi$. Then $\Phi$ is a critical point of the Hitchin Functional when it is restricted to the cohomology class of $\Phi$. Conversely, let $\Phi$ be a critical point of the Hitchin Functional on the cohomology class that is also a positive three-form. Then a seven-manifold M equipped with $\Phi$ defines a G$_{2}$-Metric.} 
\end{thm}

Note that the cohomology class of $\Phi$ is contained in H$^3$(M, $\mathbb{R}$), to which we know from Joyce's result that the moduli space is locally diffeomorphic. We can recover this result in the Hitchin formalism by noting that the functional has a Morse-Bott critical point. After noting this fact, the argument proceeds somewhat analogously to that of Joyce. \\

So either via Joyce's method or directly from the Hitchin Functional we may note that the moduli space is locally diffeomorphic to H$^3$(M, $\mathbb{R}$). Note that the fact that the moduli space is a smooth manifold may be deduced directly from the fact that it is locally diffeomorphic to a smooth submanifold; such a result requires no global knowledge. This result is important for determining the local structure of the moduli space of torsion-free $G_{2}$-Metrics, but additional local information may be gathered by considering a point in the moduli space corresponding to a particular $G_{2}$-Structure. This $G_{2}$-Structure is determined by a torsion-free positive three-form, so we could study the local structure of the moduli space by considering infinitesimal displacements away from the original three-form. If the original three-form is denoted $\Phi$, then these considerations amount to considering the perturbed three-form $\Phi'$. 
\begin{equation}
\Phi'=\Phi +\epsilon \xi
\end{equation}

We can then determine the deformations of the coassociative four-form and the metric by replacing $\Phi$ with $\Phi '$. The important detail to keep in mind is that, in order to define a torsion-free G$_{2}$-Structure, the torsion of the perturbed three-form must also vanish, which imposes an additional condition. It might also seem that we need to be concerned about the factor $\epsilon$; however, we can always chose $\epsilon$ to be small enough such that the perturbed three-form is positive since the set of positive three-forms is an open sub-bundle of $\Lambda^3 T^{*}M$. To first order, the constraint that the deformation three-form $\Phi '$ be torsion-free is equivalent to the constraint that the perturbation $\xi$ be closed and co-closed. In the case that the manifold is compact this is equivalent to $\xi$ being harmonic. \\

Taking into account these constraints one can write down an expansion in $\epsilon$ for the perturbed metric as well. This procedure for the coassociative four-form and metric has been carried out to low orders in $\epsilon$ \cite{Karigiannis, gy}. \\

We can use Joyce's result to deduce additional structure for the local moduli space, in particular to define a local notion of the metric. Using the fact that the moduli space is locally diffeomorphic to H$^3$(M, $\mathbb{R})$ we can identify an affine coordinate system on H$^3$(M, $\mathbb{R})$ by expanding an element in a basis and taking the affine coordinates as the expansion coefficients. This coordinate system corresponds to a flat connection. Based on these considerations, it follows that the moduli space of torsion-free $G_{2}$-Structures admits a Hessian metric \cite{Grigorian}. 
\begin{definition} Hessian Manifold and Potential. \textnormal{Let M be a smooth manifold with a flat, torsion-free connection. A Riemannian Metric g is called \textit{Hessian} and (M, g) is a \textit{Hessian Manifold} if g may be expressed locally in the following form.} 

\begin{equation}
g_{ab}=\frac{\partial^2 H}{\partial x^{a}\partial x^{b}}
\end{equation}

\textnormal{Here, x$^{a}$ is an affine coordinate system. In this case, the function H is called the \textit{Hessian Potential}.}

\end{definition}

This relationship between the Hessian Metric and Potential looks similar to that of the K{\"a}hler metric and potential. In a sense, the Hessian structure may be thought of as the analogue of K{\"a}hler structure for a real manifold. More precisely, the complexification of a Hessian Manifold is K{\"a}hler \cite{Grigorian}. As a consequence, the specification of a Hessian Potential on a local region of the moduli space determines a metric on this region.  The natural choice for the Hessian Potential is given directly in terms of the Hitchin Functional. \\

In the literature, there is some ambiguity regarding the functional dependence of the Hessian Potential on the Hitchin Functional F. We will follow the physics convention and take the potential to be proportional to the logarithm of the Hitchin Functional. 
\begin{equation}
H=-3\log{F}
\end{equation}

 The basic reason for this choice is in analogy to K{\"a}hler geometry, for in the physics context we will see that there is a natural complexification for the moduli space, which is K{\"a}hler following from the fact that the G$_{2}$ moduli space is Hessian. In any case, we can differentiate the potential to find the local metric. 
\begin{equation} 
G_{ab}=\frac{1}{F} \int_{M} \Phi_{a} \wedge \psi_{b}
\end{equation}

\subsection{Global Aspects and Future Directions}

It is important to keep in mind that all of the results discussed are only valid locally on the moduli space of torsion free $G_{2}$-Structures. This restricted domain of validity follows from the fact that the moduli space is only locally diffeomorphic to H$^3$(M, $\mathbb{R})$. Moreover, the deformation analysis of the $G_{2}$-Metric is only well-understood to low orders in $\epsilon$. In general such results are not known to be valid for arbitrary values of $\epsilon$, and calculating higher order perturbations in $\epsilon$ is computationally involved. All that is really known generically about the global nature of the moduli space is that it is a smooth manifold with dimension $b^3(M)$, which can be deduced directly from local properties\\

In the case of the Calabi-Yau moduli space, Tian and Todorov were able to extend the local results to the global results \cite{Tian,Todorov}, and locally, much of the analysis in the Calabi-Yau case is similar to the $G_{2}$ case. The difficulty in extending the results rests primarily in the non-linearity of the torsion-vanishing condition on the $G_{2}$ three-form. Nevertheless, understanding the global structure of the $G_{2}$ moduli space is of crucial importance. In particular, it would be advantageous to know if the map $\pi$: M $\rightarrow$ H$^3$(M, $\mathbb{R})$ is injective, and to determine its image \cite{JoyceB}.  \\

However, there has been some recent progress with regards to the torsion-free G$_{2}$ moduli space. Much of the success has had to do with sophisticated invariants which can be used to distinguish homeomorphic and diffeomorphic G$_{2}$-Manifolds. For instance, Crowley-Nordstr{\"o}m identified a $\mathbb{Z}_{48}$-valued homotopy invariant $\nu$, that together with another homotopy invariant $\xi$ is able to determine a G$_{2}$-Structure of a two-connected manifold up to homotopy and diffeomorphism \cite{cn}. Building off these results, Crowley-Goette-Nordstr{\"o}m were able to identify seven-manifolds for which the moduli space of torsion-free G$_{2}$-Structures is disconnected, and they found examples of homeomorphic but not diffeomorphic G$_{2}$-Manifolds \cite{cgn}.  \\

The moduli space we have considered in this section corresponds to that of compact manifolds that admit torsion-free $G_{2}$-Structures and the space by assumption has fixed topology. The moduli space could be generalized in several ways to consider additional aspects relevant to mathematics and physics. Perhaps most importantly for physics, one could consider a moduli space corresponding to $G_{2}$-Spaces with prescribed singularities. Understanding such a moduli space is of paramount importance for M-Theory. 

\section{The Supergravity Approximation}

Despite the fact that the action for M-Theory is not known (if it can be formulated in terms of an action at all), the physics of M-Theory may be studied in the low energy limit because this limit coincides with $\mathcal{N}=1$ eleven-dimensional supergravity. Studying M-Theory in this limit is commonly referred to as the supergravity approximation. It is also worth noting that eleven-dimensional supergravity is special in its own right as the maximal dimensional supergravity theory to exist in nature (not admitting particles of spin greater than 2) \cite{Witten1}. This distinguishing feature of the eleven-dimensional supergravity theory adds weight to the significance of M-Theory through its relation via the supergravity approximation. \\

\subsection{Fundamentals of Eleven-Dimensional $\mathcal{N}=1$ Supergravity}

The advantage of studying M-Theory in the supergravity approximation is that the supergravity action is known and it allows us to directly study how the topology of the M-Theory compactification space determines specific features of the action. Before doing so, it is worth noting that the properties of a compactification on a G$_{2}$-Manifold are consistent with the mathematical requirements of an eleven-dimensional supergravity theory. Such a theory is specified by a smooth Lorentzian eleven-manifold with metric g$_{ab}$ and three-form field C$_{abc}$. Dynamics of the C-Field following from the supergravity action are usually characterized in terms of the four-form flux $\textnormal{G}=\textnormal{dC}$. Topologically, the Lorentzian Manifold M is assumed to be spin and its first Pontryagin class must satisfy the following cohomological condition. 
\begin{equation}
\left[\frac{G}{2\pi}\right] -\frac{p_{1}(M)}{4} \in H^4(M, \mathbb{Z})
\end{equation}

As a preliminary step to studying eleven-dimensional $\mathcal{N}=1$ supergravity on spaces of the form $M_{4} \times M_{7}$ where $M_{4}$ is non-compact, Lorentz, and flat and where $M_{7}$ is a (compact) G$_{2}$-Manifold, we need to ensure that such a spacetime satisfies the aforementioned criteria. As discussed previously, we know that G$_{2}$-Manifolds are spin and moreover that simply-connected examples are associated with a preferred spin structure. We also know that in the compact case the first Pontryagin class of a G$_{2}$-Manifold is necessarily non-zero. Regardless as to what it is, the supergravity theory is well-posed as long as the four-form flux G is consistent with it as in equation 5.1. These considerations regarding the G$_{2}$-Manifold generalize to the entire spacetime $M_{4} \times M_{7}$, so such a background is consistent with the supergravity requirements. \\ 

Focusing now on the content of the physical theory, the relevant massless bosonic fields are the eleven-dimensional metric $g_{ab}$ and the three-form $C_{abc}$ and the massless fermionic field is the eleven-dimensional gravitino $\Psi_{a}$. These fields transform in the following representations of SO(9) \cite{gjky}. 
\begin{itemize}
\item Metric: Traceless Symmetric Representation \textbf{44}
\item Three-Form Field: Anti-Symmetric Three-Tensor Representation \textbf{84}
\item Gravitino: Spinorial Representation $\textbf{128}_{s}$
\end{itemize} 

The dynamics of these fields follow from the eleven-dimensional $\mathcal{N}=1$ supergravity action, given in terms of the eleven-dimensional Hodge star, gamma matrices, and Ricci Curvature R. The complete action may be expressed in the form $S_{11D}=S_{1}+S_{2}+S_{3}+S_{4}$ where the summands denote the kinetic terms, interaction terms, Chern-Simons terms, and four-fermion interactions, respectively. These individual pieces may be written down in the following ways \cite{gjky}.
\begin{align}
& S_{1}=\frac{1}{2\kappa_{11D}^2}\int_{M} \star R -\frac{1}{2} G \wedge \star G -\star i\bar{\psi_{a}}\Gamma^{abc}\nabla_{b}\psi_{c}\\
& S_{2}=-\frac{1}{192\kappa_{11D}^2} \int_{M} \star\bar{\psi_{a}}\Gamma^{abcdef}\psi_{b}G_{cdef} -\frac{1}{2\kappa_{11}^2} \int_{M} G \wedge \star 3\bar{\psi_{[a}}\Gamma_{bc}\psi_{d]}  \\
& S_{3}=-\frac{1}{12\kappa_{11D}^2} \int_{M} G \wedge G \wedge C
\end{align}

The coupling constant $\kappa_{11D}$ is related to the eleven-dimensional Newtonian constant, or alternatively the Planck length.

\subsection{Kaluza-Klein Reduction on G$_{2}$-Manifolds}

Given an eleven-dimensional $\mathcal{N}=1$ supergravity theory, we can perform a Kaluza-Klein reduction of the fields to understand the massless spectrum of the four-dimensional effective theory. This reduction will utilize the compactification ansatz and a corresponding metric ansatz to write the fields in terms of mode expansions. Using these mode expansions, we can then write down an action for the effective four-dimensional theory given in terms of quantities determined by the M-Theory compactification space. We will use the typical compactification ansatz and most generally, we could consider a warped metric.
\begin{equation}
g_{ab}=\Delta^{-1}(y)(\eta_{\mu\nu}dx^{\mu}dx^{\nu}+g_{\mu\nu}'(y)dy^{\mu}dy^{\nu})
\end{equation}

Aspects of such warped compactifications have been studied in the literature \cite{bb, bw}. However, if the compactification space is compact and we want $\mathcal{N}=1$ supersymmetry in the effective theory, then the warp factor and G-Flux must vanish \cite{as}. This vanishing of the G-Flux justifies the assumption (at least in the compact case) that we made in the introduction leading up to the existence of a covariantly-constant spinor. Since the warp factor vanishes in this case we may use a simpler metric ansatz. 
\begin{equation}
g_{ab}=\eta_{\mu\nu}dx^{\mu}dx^{\nu}+g_{\mu\nu}'(y)dy^{\mu}dy^{\nu}
\end{equation}

 We can then perform the mode expansions for the relevant fields using this metric ansatz and the cohomological data of the compactification space. Concerning the mode expansion for the metric, the important detail to keep in mind is that to linear order, the metric perturbation is constrained to satisfy the vacuum Einstein's Equations. Ultimately, since our space is compact this requirement is equivalent to the first order three-form deformation of the moduli space being harmonic. The massless four-dimensional $\mathcal{N}=1$ multiplet associated with the Kaluza-Klein reduction of the eleven-dimensional metric is the four-dimensional gravity multiplet. For future use let S$^{i}$ characterize a point in this moduli space. \\
 
 Likewise, we can consider the mode expansion of the C field. We can expand it in terms of harmonic two-forms $\omega$ and three-forms $\rho$.  
 \begin{equation}
 C=A^{I} \wedge \omega_{I} + P^{i}\rho_{i}
 \end{equation}
 
 Here the index I runs from 1 to b$_{2}$(M) and the index i from 1 to b$_{3}$(M). As a consequence, associated with the C field are massless modes in the form of b$_{2}$(M) vectors A$_{I}$ and b$_{3}$(M) pseudoscalars P$_{i}$. \\
 
 Lastly, we consider the dimensional reduction of the gravitino, which will be associated with the massless fermionic modes. For the eleven-dimensional gravitino, the zero modes correspond to those of the seven-dimensional Dirac and Rarita-Schwinger Operators. Analogous reduction leads to the four-dimensional gravitino, b$_{3}$(M) spinor fields, and b$_{2}$(M) gauginos. So altogether, the massless four-dimensional $\mathcal{N}=1$ multiplets are one gravity multiplet, b$_{3}$(M) chiral multiplets, and b$_{2}$(M) vector multiplets. \\

Plugging in the mode expansions into the eleven-dimensional supergravity action, we can directly write down the four-dimensional bosonic supergravity action \cite{gjky}. 
\begin{align}
& S_{4D}^{\textnormal{bos}}=\frac{1}{\kappa_{4D}^2} \int \star_{4} R + \frac{\kappa_{IJk}}{2}S^{k}F^{I} \wedge \star_{4} F^{J} -P^{k}F^{I} \wedge F^{J} \\
& -\frac{1}{2\lambda} \int \rho_{i} \wedge \star_{7} \rho_{j}(dP^{i} \wedge \star_{4} dP^{j} +dS^{i} \wedge \star_{4} dS^{j})
\end{align}

In this expression, the constants $\kappa_{4D}$ and $\lambda$ are related to a moduli-dependent volume factor, which is related to the Hitchin Functional, and F is shorthand for $3\bar{\psi_{[a}}\Gamma_{bc}\psi_{d]}$ The $\kappa_{IJk}$ are determine by the following integral.
\begin{equation}
\kappa_{IJk}=\int \omega_{I} \wedge \omega_{J} \wedge \rho_{k}
\end{equation}

Through the process of inserting the mode expansions for the fields, the action is not yet in the form typical for that of four-dimensional $\mathcal{N}=1$ supergravity \cite{wb}. Such an action is given in terms of the vector and chiral multiplets whose existence we inferred from dimensional reduction. Present in the standard action is the K{\"a}hler Potential, associated with the K{\"a}hler structure of the target space, and gauge kinetic terms, so we want to figure out how to express these in terms of the action derived via dimensional reduction. A preliminary step is to note the complex coordinates on the K{\"a}hler target space. To first order these are dictated by membrane instanton corrections to the $\mathcal{N}=1$ superpotential under the assumption of holomorphy of the superpotential \cite{hm}. The complex coordinates are then the components of the chiral field multiplets, given in terms of the coordinates $P^{i}$ from the dimensional reduction of the C-Field and the coordinates S$^{i}$ from that of the eleven-dimensional metric. 
\begin{equation}
\phi^{i}=-P^{i}+iS^{i}    
\end{equation}

Then we can rewrite the bosonic action and read off the forms of the K{\"a}hler Potential and Gauge Kinetic Functions. Note the relationship between the K{\"a}hler Potential presented here and the complexified Hessian potential introduced during the moduli space discussion. 
\begin{align}
& K(\phi,\bar{\phi})=-3\log{\frac{1}{7}\int_{N} \phi \wedge \star\phi} \\
& f_{IJ}=\frac{i}{2}\sum_{k}\phi_{k}\int_{N} \omega_{I} \wedge \omega_{J} \wedge \rho_{k} 
\end{align}

In these expressions the integration is over the G$_{2}$-Manifold N.\\

In the event that we have non-vanishing G-Flux, we can model this effect on the action by including a flux-induced superpotential \cite{bw}. In terms of its effect on the action, the terms associated with this superpotential enter quadratically in the bosonic action but linearly in the fermionic action. As a result, it is easiest to directly read off the terms of this superpotential via the fermionic terms by means of dimensional reduction, analogously to the procedure for the K{\"a}hler Potential and Gauge Kinetic Functions. Doing so, it is given by 
\begin{equation}
W(\phi)=\frac{1}{4}\int_{N} G \wedge \left (-\frac{1}{2}C+i\phi\right)
\end{equation}

For much of this discussion, the connection to the mathematical structure of the compactification space has been indirect, but present. Note for instance that the Betti numbers of the compactification space determine the number of vector and chiral multiplets in the effective theory. \\

So far, this procedure has made no assumptions about the geometrical nature of the seven-dimensional compactification space, so in that regard was completely general. However, we are particularly interested in M-Theory compactifications so an interesting question is whether there are any special aspects that arise in the effective supergravity theory as a result of compactifying on a space of G$_{2}$-Holonomy. \\

To the end of studying the effective supergravity action, the twisted connected sum G$_{2}$-Manifolds have been most amenable to study \cite{gjky}. The basic reason for their favorability follows from the fact that in a certain limit, their metrics and cohomology are determined to leading order by those of the Calabi-Yau summands. Physically, this limit is roughly analogous to the large volume limit of compactifications of the Type II strings on a Calabi-Yau. As is generally true the Betti numbers determine the spectrum, but in the spectrum two chiral fields are of particular interest. One such field $\nu$ parameterizes the volume of the G$_{2}$-Manifold, and another $\chi$ parameterizes this limit in which the G$_{2}$-Manifold is determined by the summands. Given these interpretations of the fields, we can evaluate the K{\"a}hler Potential directly in these appropriate limits. 
\begin{equation}
K=-\log{((\nu+\bar{\nu})^4(\chi+\bar{\chi})^3)}
\end{equation}

This knowledge of the effective theory makes it possible to study various low energy aspects of M-Theory on G$_{2}$-Manifolds, but specifically those of twisted connected sum type, and only in the appropriate limit. Although realistic M-Theory compactification spaces are necessarily singular, having this explicit knowledge of the action in the supergravity limit, even in the case that the space is smooth, is helpful for determining whether the space can accommodate appropriate singularities. 

\section{M-Theory, Dualities, and G$_{2}$-Manifolds}

In the previous section we saw how to characterize the physics of M-Theory in the low energy limit through its relationship to eleven-dimensional $\mathcal{N}=1$ supergravity. However, it is obviously fundamental to study M-Theory outside of the supergravity limit and understand how the geometrical aspects of the M-Theory compactification space determine the effective physics. Even though the complete formulation of the theory is not known, we can still study aspects of M-Theory through its dualities with the superstring theories. For the effective theory to be realistic, we need to be able to achieve the appropriate non-Abelian gauge symmetry and chiral matter. Crucially, in order to achieve either of these goals the compactification space must be singular. We will study these singularities in the next section, instead first focusing on aspects of M-Theory on G$_{2}$-Manifolds that are interesting in and of themselves, and also in regards to achieving standard model phenomenology. Namely, we will study topological transitions in M-Theory and notions of mirror symmetries involving G$_{2}$-Manifolds.  

\subsection{Topological Transitions in M-Theory}

A topological transition corresponds to a change in topology of a space as it approaches a singular limit. These are commonly studied in the context of local models of M-Theory compactification spaces (complete and non-compact G$_{2}$-Manifolds) since the singular limit corresponding to these transitions may lead to enhanced gauge symmetry. Recall that the non-compact examples studied in depth were either asymptotically conical or asymptotically locally conical. As a consequence of these geometries, these spaces have singular limits associated with conical singularities. For instance, we can express the Bryant-Salamon spaces in the following form, where the space C is contractible \cite{ag}.
\begin{equation}
M \cong B \times C 
\end{equation} 

With M expressed in this form, the singularity corresponds to the limit in which Vol(B) goes to zero. These limits can potentially correspond to topological transitions. These types of topological transitions are also familiar in the context of Calabi-Yau Manifolds in string theory, where the two examples are the conifold transition and the flop transition. The flop is a real phase transition in which a three-cycle is replaced by a two-cycle \cite{agm} and the conifold is a smooth transition in which a three-cycle is replaced by a distinct three-cycle \cite{gms}. A topological transition is smooth if the quantum theory is not subject to any singular behavior, but a phase transition experiences discontinuities of parameters in the quantum theory. \\

There are also two broad classes of topological transitions in the M-Theory context. In a G$_{2}$ Flop transition a three-cycle is replaced by a distinct three-cycle (somewhat analogously to the conifold transition). Such a transition may be realized on the rank three Bryant-Salamon total space \cite{amv}. 
\begin{equation}
M=S^3 \times \mathbb{R}^4
\end{equation}

On such a space the G$_{2}$ Flop transition is smooth \cite{aw}. \\

Another M-Theory transition is called the Phase Transition. As its name would suggest, it is a real phase transition \cite{aw}. It can occur on the following rank four Bryant-Salamon total space. 
\begin{equation}
\mathbb{C}P^2 \times \mathbb{R}^3
\end{equation}

One distinctive aspect of these two M-Theoretic topological transitions on these spaces concerns the quantum moduli space. In both cases the quantum moduli space has three branches, but in the case of the G$_{2}$ flop it is possible to go between the branches without going through the singularity though in the phase transition case it is necessary to go through the point at which the geometry becomes singular. \\

An interesting question concerns the relationship between topological transitions in the context of M-Theory and the superstring theories. For instance, it is a priori interesting that both the G$_{2}$ flop transition and conifold transitions involve replacing a three-cycle with another three-cycle, even though the underlying geometries are only related via special holonomy (SU(3) and G$_{2}$). In specific circumstances it is possible to relate these two transitions by breaking supersymmetry further via D-Branes. \\

Starting with type IIA on $S^3 \times \mathbb{R}^3$, we can wrap such a space-filling D6-Brane around the Special Lagrangian (the calibrated submanifold for Calabi-Yau manifolds) and then undergo a conifold transition. Doing so, we get type IIA on $S^2 \times \mathbb{R}^4$ with RR flux. This relationship is called a geometric transition and it was originally rather unexpected in the context of string theory \cite{Vafa}. However, this geometric transition may be anticipated in the M-Theory context \cite{amv}. In fact, lifting both sides to M-Theory this geometric transition is nothing more than the G$_{2}$-Flop, which is a relatively intuitive being a topological transition. \\

From this example we have seen that it is sometimes possible to relate transitions between string theory compactifications to transitions between G$_{2}$-Manifolds. These relationships are especially fruitful to study in the case of G$_{2}$-Manifolds constructed via twisted connected sum due to the intimate relationship between the G$_{2}$-Manifolds and their Calabi-Yau summands. In specific limits, the transitions between G$_{2}$-Manifolds of twisted connected sum type may be interpreted in terms of transitions of their Calabi-Yau summands \cite{chnp1}. However, it is important to keep in mind that Fano three-folds undergoing conifold transitions will generically only become weak Fano three-folds. Understanding this subtlety, it is still possible in principle to match building blocks such that the transitions of the blocks effectively signify a transition between the corresponding G$_{2}$-Manifolds. \\

Although the twisted connected sum examples are particularly amenable to study through topological transitions, we could also consider singular behavior in more general G$_{2}$-Manifolds that could lead to a transition. In the case that a G$_{2}$-Manifold admits an associative or coassociative submanifold, promising singular behavior could appear as the volume of the submanifold approaches zero. By explicitly incorporating the calibrated submanifolds, the issue of understanding the physics of these limits becomes more tractable in certain cases since they are closely related to topological defects in the context of wrapped branes \cite{hm2}. In some circumstances they are also useful for controlling calibrated two-cycles to the end of controlling particle masses.    

\subsection{Mirror Symmetry involving G$_{2}$-Manifolds}

Mirror Symmetry, as the term is typically used, refers to a non-trivial relationship between topologically distinct Calabi-Yau Manifolds used as compactification spaces for Type IIA and IIB string theories. The physical motivation for mirror symmetry follows from the correspondence between states of the Conformal Field Theories (CFTs) of the Type II string theories compactified on the distinct Calabi-Yau manifolds. Related phenomena may occur in the study of M-Theory compactified on G$_{2}$-Manifolds. One interesting possibility is the presence of a non-trivial relationship between G$_{2}$-Manifolds underlying M-Theory compactifications and Calabi-Yau Manifolds underlying string theory compactifications. However, there could also be interesting relationships between pairs of G$_{2}$-Manifolds, just as the original notion of mirror symmetry relates two Calabi-Yau Manifolds. \\

While the understanding of mirror symmetry between Calabi-Yau Manifolds from a mathematical standpoint remains an area of active research, in the context of physics it is conjectured to be a fiberwise T-duality \cite{syz}. This approach to mirror symmetry is called Strominger-Yau-Zaslow (SYZ) Mirror Symmetry and it suggests that the fibres are T$^{3}$. In other words, presumably any Calabi-Yau three-fold participating in mirror symmetry is fibered by a T$^3$. \\

Accepting this notion of SYZ Mirror Symmetry, we can ask what implications it has for M-Theory. Namely, as M-Theory is related to the superstring theories, then if the string theory compactification space is fibered by a T$^3$, it stands to question whether that implies anything about the corresponding G$_{2}$-Manifold in M-Theory. \\

We can answer this question by utilizing the duality between M-Theory compactified on a K3 surface and Heterotic String Theory on a T$^3$ \cite{gyz}. Considering a Calabi-Yau Manifold with a T$^3$ fibration, as in SYZ mirror symmetry, we can consider heterotic string theory on the fibers and utilize the duality with M-Theory on a K3 to suggest that G$_{2}$-Manifolds dual to Calabi-Yau three-folds with T$^{3}$ fibrations themselves have K3 fibrations. \\

Besides the duality of M-Theory with heterotic string theory, we could consider the duality of the former with type IIB theory with RR flux. Considerations regarding the moduli space of an M-theory five-brane wrapped around a four-cycle suggest that the G$_{2}$-Manifold should admit a T$^4$ fibration \cite{gyz}. Considering both of these dualities, it stands to reason that G$_{2}$-Manifolds admitting string theory duals should be fibered by coassociative submanifolds \cite{gyz}. \\

These fibration arguments most directly suggest a relationship between Calabi-Yau Manifolds and G$_{2}$-Manifolds, though indirectly could suggest a genuine relationship between G$_{2}$-Manifolds. More directly however, several notions of "exceptional mirror symmetry" exist between pairs of G$_{2}$-Manifolds, both for Joyce orbifold resolutions and twisted connected sums. \\

In the original context of mirror symmetry as a relationship between Calabi-Yau Manifolds, the physical impetus suggesting a mathematical relationship was rooted in the underlying CFTs. We could also approach exceptional mirror symmetry in this manner to the end of identifying a mirror symmetry-type relationship between distinct G$_{2}$-Manifolds \cite{sv}. Specializing to Joyce orbifold resolutions, to the end of producing the same CFT starting from distinct spaces, the CFT was not sensitive to the Betti numbers b$_{2}$(M) or b$_{3}$(M) individually; however, the sum b$_{2}$(M)+b$_{3}$(M) was the same in each case. Interestingly enough, looking at a plot of Betti numbers achieved via the Joyce orbifold resolutions, the pairs (b$_{2}$(M), b$_{3}$(M)) are commonly on diagonal lines of constant sum b$_{2}$(M)+b$_{3}$(M) \cite{JoyceB}. This CFT-based reasoning, as well as this mathematical curiosity involving the Betti numbers, is suggestive of a mathematical relationship between G$_{2}$-Manifolds with the same sums b$_{2}$(M)+b$_{3}$(M). However, it is possible that such reasoning holds only for G$_{2}$-Manifolds constructed via orbifold resolutions. \\

Exceptional Mirror Symmetry also has been considered on Joyce orbifold resolutions in the context of duality between Type II string theories \cite{Acharya44}. In this context these theories are compactified on specific G$_{2}$-Manifolds of Joyce Resolution type and then studied using SYZ mirror symmetry. In the case of a duality between Type IIA theory on one manifold and Type IIB on another, these considerations suggest that the G$_{2}$-Manifolds should be fibered by T$^3$s. Alternatively, for a mirror symmetry to occur between the same Type II theory on different G$_{2}$-Manifolds, these manifolds should be T$^4$ fibrations.  \\

With the advent of twisted connected sum G$_{2}$-Manifolds, it became possible to test whether these notions of exceptional mirror symmetry were limited to the orbifold case or if they were more generic features of holonomy G$_{2}$. Type II mirror symmetries have been explored on these twisted connected sum backgrounds \cite{bdz}. One key result is that, like Joyce Orbifold Resolutions, twisted connected sums may exhibit both the mirror symmetries associated with the same Type II theory, in which case they are fibered by T$^4$, and the mirror symmetries between Type IIA and Type IIB, in which case they are fibered by T$^3$. This result, at least, is then not a consequence of the specific features of the Joyce construction. \\

\section{Achieving Standard Model Phenomenology}

As we have seen, one regime in which we can study M-Theory is in the low energy limit, which corresponds to eleven-dimensional $\mathcal{N}=1$ supergravity. To the end of understanding the relationship of M-Theory to the observable world, we also want to know in what regime it corresponds to standard model phenomenology. The considerations leading up to G$_{2}$-Holonomy imply that the effective theory will have $\mathcal{N}=1$ supersymmetry, but more specifically we want to be able to achieve four-dimensional $\mathcal{N}=1$ Super Yang-Mills and at some energy scale break supersymmetry to the standard model gauge group SU(3) $\times$ SU(2) $\times$ U(1). In order to achieve both this non-Abelian gauge symmetry and chiral fermions, we will need to introduce singularities into the compactification space. We focus on motivating these specific types of singularities as opposed to studying four-dimensional $\mathcal{N}=1$ Super Yang-Mills and the onset of quantum aspects. \\

The fact that realistic M-Theory compactifications must be singular is also interesting in relation to the supergravity limit. After all, supergravity is only well-defined on a smooth space, so somewhere in the process of taking the supergravity limit of M-Theory on a singular space the singularities must resolve themselves. The exact mechanism for this resolution of singularities is not well-understood, and would probably require knowledge of the underlying microscopic degrees of freedom of M-Theory. It is important to keep in mind the fundamental role that supergravity plays in the study of M-Theory compactifications. For instance, it was reasoning from supergravity that suggested that we should be interested in G$_{2}$-Holonomy in the first place, and such reasoning also implies that we should look at singular spaces as we will do in this section.  

\subsection{Non-Abelian Gauge Symmetry}

As a consequence of the fact that compact G$_{2}$-Manifolds exhibit no continuous symmetries, the gauge fields are determined by the Kaluza-Klein reduction of the C-Field \cite{Witten2}. Considering this dimensional reduction, compactifying M-Theory on a smooth G$_{2}$-Manifold leads to the gauge group U(1)$^{b_{2}(M)}$ \cite{hm2}, which is necessarily Abelian. However, on phenomenological grounds we need to be able to achieve non-Abelian gauge symmetry. Fortunately, singularities are able to enhance the gauge symmetry to non-Abelian gauge groups such that we have a hope of making contact with standard model physics. Understanding the role of singularities towards achieving physically-realistic theories, it would be very useful to be able to categorize these possible singularities that are physically-realistic. Unfortunately, such a comprehensive list of singularities is not known (even in the Calabi-Yau case) so the current approach focuses on studying specific promising examples \cite{ag}. \\

We can motivate promising examples of singularities by considering the dualities of M-Theory with the superstring theories. One particularly useful duality is that between M-Theory on a K3 surface and Heterotic String Theory on a T$^{3}$. Recall that we also made use of this duality in relating SYZ mirror symmetry to the mirror symmetry of Gukov-Yau-Zaslow. The advantage of using this particular duality follows from the fact that the moduli space of SU(2) metrics on a K3 surface is known, locally at least, to be of the following form \cite{ag}. 
\begin{equation}
M=\mathbb{R}^{+} \times \frac{SO(3,19)}{SO(3) \times SO(19)}
\end{equation}

Moreover, a toroidal compactification of heterotic string theory preserves supersymmetry. Noting that supersymmetry is then automatically preserved on the heterotic side of the duality, it must also be preserved for M-Theory compactified anywhere on the K3 moduli space, including in the singular regions. This preservation of supersymmetry on the M-Theory side constrains the behavior of the K3 singularities. Specifically, we can generically write the orbifold singularities of the K3 in the following form. \\

\begin{equation}
N=\mathbb{C}^2 \slash ~\Gamma
\end{equation}

Here, $\Gamma$ is a finite subgroup of SO(4). Then for supersymmetry not to be broken in the M-Theory compactification, the finite group $\Gamma$ must actually be a finite subgroup of SU(2). The corresponding finite subgroups are known and the singularities fall under the designation of A-D-E singularities \cite{Polchinski}. 
\begin{definition} A-D-E Singularity. \textnormal{An \textit{A-D-E Singularity} is an orbifold singularity of the form $\mathbb{C}^2 /\Gamma$ where $\Gamma$ is the finite subgroup of SU(2) associated with one of the simply-laced Lie Algebras A$_{n}$, D$_{k}$, E$_{6}$, E$_{7}$, or E$_{8}$} 
\end{definition}

Intuitively, such singularities are associated with the collapse of two-spheres to zero volume. In this limit, the duality with the heterotic string implies that the gauge symmetry is enhanced to the finite subgroup of SU(2) associated with the singularity \cite{Acharya1}.  \\

Using this duality of M-Theory on a K3 with heterotic string theory on a T$^{3}$, we have inferred that A-D-E type singularities are relevant for achieving non-Abelian gauge symmetry. However, we came to this understanding through the use of a particular duality between seven-dimensional theories. It is not a priori obvious that, in its current context, the specialization to A-D-E singularities is particularly relevant for four-dimensional gauge enhancement. \\

Fortunately, there is reason to expect that A-D-E singularities are still relevant for achieving non-Abelian gauge symmetry outside of the context of this particular duality. For instance, we can imagine embedding A-D-E type singularities into G$_{2}$-Manifolds and then considering the four-dimensional theory arising from compactification. In this context the singularities would be codimension four. In the large volume limit for the G$_{2}$-Manifold (such that we can use the supergravity approximation), we can expect that the four-dimensional gauge symmetry associated with the seven-dimensional Super Yang-Mills Theory is non-Abelian and is determined by the A-D-E singularity. In this regime at least, we can expect to achieve a four-dimensional non-Abelian gauge group \cite{ag}. \\

The use of the duality between M-Theory on a K3 surface and heterotic string theory on a T$^3$ is one way to identify the relevance of A-D-E singularities to non-Abelian gauge symmetry, although there are many other avenues that also are suggestive of A-D-E singularities. For instance, another useful duality whose considerations lead to the same conclusion is that with Type IIA string theory and intersecting D6-Branes \cite{Witten2}. On the Type II with Intersecting D6-Branes side of the duality, the gauge enhancement is known at the points of intersection and is also suggestive of A-D-E type singularities in the M-Theory context.  

\subsection{Fermionic Matter}

In addition to the Abelian gauge group, we also saw that Kaluza-Klein reduction on a smooth G$_{2}$-Manifold leads to b$_{3}$(M) massless neutral complex scalar fields. "Neutral" here means that they are not charged under the gauge group, but what we really want is charged chiral matter. Charged chiral matter is phenomenologically desirable because it is massless so long as the gauge symmetry it is charged under remains unbroken.\\ 

The question now is what classes of singularities lead to chiral fermions. Similar to the case of the singularities associated with non-Abelian gauge symmetry, we do not have an adequate classification of possible singularities so we need to look at specific examples. As we want chiral matter in four dimensions, we are motivated to consider singularities of maximal codimension in the G$_{2}$-Manifolds, or in other words, isolated singularities \cite{ag}. The known isolated singularities of G$_{2}$-Manifolds are conical singularities, and fortunately, such singularities are conducive to the existence of chiral matter in four dimensions. \\

Although there are many specific types of conical singularities, we can motivate interest in this family of singularities broadly by considering the (local) anomalies of the low-energy effective action corresponding to gauge transformations. We specify the geometry, including the singularity structure of the theory, and then provided that the theory is consistent we expect the anomaly associated with variations of terms in the bulk to vanish. Such anomaly cancellation is indicative of the presence of chiral fermions. Starting with a G$_{2}$-Manifold with a conical singularity, with metric modeled by that of equation 3.1, the anomaly may be shown to vanish \cite{Witten2}. More importantly, this reasoning still holds up under certain conditions if the G$_{2}$-Manifold incorporates A-D-E singularities. These singularities determine the groups under which the chiral fermions are charged: Chiral fermions appear in the spectrum when the A-D-E singularity locus passes through points corresponding to conical singularities \cite{achw}. \\

Having motivated interest in conical singularities broadly via considerations from anomaly cancellation, we can study specific examples. The difficulty with studying these singularities in relation to M-Theory follows from the fact that either these singularities are easy to motivate via dualities and difficult to study mathematically, or the singularities are understood mathematically and difficult to motivate physically. \\

Concerning the former class of physically-motivated conical singularities, an example that has been studied using different perspectives has isolated singularities of the form M/U(1) where M is a hyperk{\"a}hler eight-manifold. These singularities were first motivated by duality with the heterotic string \cite{achw}, and then studied and generalized using the Type IIA picture. Specifically, in this latter picture the M-Theory picture may be studied using duality with Type IIA with stacks of intersecting D6-Branes \cite{bb}. The singularities in these cases are unfoldings of hyperk{\"a}ler quotients. \\

Examples of conical singularities that are well-understood mathematically include those associated with $\mathbb{C}$P$^3$, SU(3)/(U(1) $\times$ U(1)), and S$^3$ $\times$ S$^3$ \cite{aw}. Note that some examples of these spaces were discussed in the context of topological transitions, namely, the resolutions are smooth and AC. In addition there is a family of examples of the form M/U(1) where the U(1) symmetry rotates the complex structures of the hyperk{\"a}ler space \cite{achw}. In this case the metric is well-understood but the physics is less tractable. 

\section{Conclusions}

Much progress has been made in the past couple decades towards constructing compact and non-compact G$_{2}$-Manifolds and studying M-Theory compactified on these spaces. However, much remains to be done. Although we have many examples of G$_{2}$-Manifolds, most arise from a relatively small number of constructions. As a result, it is not clear how representative these spaces are of G$_{2}$-Manifolds in general. New constructions would be helpful towards answering this question, but even more significant would be an existence proof analogous to Yau's proof of the Calabi Conjecture. The situation with G$_{2}$-Manifolds is not as naturally amenable to complex structure as the Calabi-Yau case, though a general existence result of any form would revolutionize the study of G$_{2}$-Manifolds. \\

In the context of realistic M-Theory compactifications, constructing singular spaces is of considerable importance. All of the study of M-Theory near singularities has relied on local models of singularities but practically-speaking, these singularities still need to be embeddable in (compact) G$_{2}$-Manifolds. Given examples of such spaces, M-Theory could be studied compactified directly on them and we could verify the validity of the reasoning drawn from the local models. All things considered, even though the mathematical problems surrounding G2-Holonomy are formidable, we have made a lot of progress towards studying M-Theory compactified on these smooth spaces, and more recently towards M-Theory on singular spaces, but there remains a lot to discover. \\

\noindent \textbf{Acknowledgments.} 

I would like to thank Andreas Brandhuber, Alex Kinsella, Sergio Hernandez Cuenca, Wayne Weng, Yaodong Li, Aleksei Khindanov, and in particular my advisor David Morrison for helpful discussions. I would also like to thank the Simons Collaboration on Special Holonomy in Geometry, Analysis, and Physics for helping me attend several recent collaboration meetings that have been instrumental to this review. Lastly, I would like to thank Jessica Li for support and encouragement. 


\end{document}